\documentstyle[a4,epsf,12pt]{article}
\begin{document}
%
%
%\title{Anomalies and Charge Quantisation Guiding Extensions of
%the Standard Model}
\title{
\begin{flushright}
\normalsize{GUTPA/95/01/2}
\end{flushright}
Extending the Standard Model Using Charge Quantisation Rules}
\author{ {\bf C.D.Froggatt, D.J.Smith} \\ {\em Department of Physics and
Astronomy,} \\ {\em University of Glasgow} \\ {\bf H.B. Nielsen} \\ {\em Niels
Bohr Institute,} \\ {\em Copenhagen} }
\date{}
\maketitle
\begin{abstract}
We examine extensions of the Standard Model (SM),
basing our assumptions
%trying to base our assumptions
on what has already been observed;
%We investigate where we are led if we seek
%the most obvious extensions of the SM in the sense that
we don't consider
anything fundamentally different,
such as grand unification or supersymmetry,
which is not directly suggested by the SM itself.
We concentrate on the possibility of additional
low mass fermions (relative to the Planck mass)
and search for combinations of representations which don't
produce any gauge anomalies.
%We also use
Generalisations of the SM weak hypercharge quantisation rule
%observed in the Standard Model
are used to specify the weak hypercharge,
modulo 2, for any given representation
of the non-abelian part of the gauge group.
%We put
Strong experimental constraints are put
on our models, by using the
renormalisation group equations to
%estimate fixed point masses for the fermions
obtain upper limits on fermion masses
%in our models
and to check that there is no $U(1)$ Landau pole
below the Planck scale.
%This is required since we are assuming
%a desert up to the Planck scale. In
Our most promising model contains
%we show that
a fourth generation of quarks without leptons
%is possible
and can soon be tested experimentally.
\end{abstract}
\newpage
\section{Introduction}
\label{Intro}
%
%********************
%\input{su5int}
Over the years there have been numerous attempts at extending the
Standard Model
(SM). Some of these models have been proposed with the purpose of explaining
some particular feature of the SM\@. For example, Grand Unified Theories (GUTs)
`explain' the convergence of coupling constants at some energy as a
manifestation of a single fundamental unified interaction. Other models
such as Supersymmetry (SUSY) have been proposed for mainly aesthetic reasons;
SUSY introduces a symmetry between bosons and fermions.
But so far none of these attempts has been entirely successful, although SUSY
GUTs are phenomenologically consistent with the unification of the SM gauge
coupling constants and do not suffer from the technical gauge hierarchy
problem.

Another approach to extending the SM is to look at the SM itself and look for
distinctive features which could be generalised or assumed to hold in an
extended theory. The SM has been so successful that, within our experimental
and calculational accuracy, it has proved to be a perfect description of nature
(except for the gravitational interaction).
So we have good reason to say that taking guidance from the SM is akin to
``listening to God".

Having accepted this point of view we must now try to interpret the message
of the SM\@. By this we mean that we must look for fundamental features in the
SM which could distinguish it from similar and, without experimental evidence,
equally plausible models. We propose that one such feature is charge
quantisation. This can be expressed as
\begin{equation}
\label{SMchqu}
\frac{y}{2}+\frac{1}{2}\rm{``duality"}+\frac{1}{3}\rm{``triality"}
	\equiv 0\pmod{1}
\end{equation}
where $y$ is the conventional weak hypercharge. The duality has value 1 if
the representation is an $SU(2)$ doublet ({\bf 2}) and 0 if it is
an $SU(2)$ singlet ({\bf 1}).
The triality has value 1 if the representation is an $SU(3)$ triplet
({\bf 3}), 0 if
it is an $SU(3)$ singlet ({\bf 1}), and -1 if it is an $SU(3)$ anti-triplet
(${\bf \overline{3}}$). In general
we can define N-ality to be the number of N-plet representations of $SU(N)$
which must be combined to give the representation of $SU(N)$. In particular
N-ality has value 1 if a representation is an $SU(N)$
N-plet (${\bf N}$), 0 if it is an $SU(N)$ singlet ({\bf 1}), and -1
if it is an $SU(N)$
anti-N-plet (${\bf \overline{N}}$). Note that in $SU(2)$ the
${\bf \overline{2}}$ representation is
equivalent to the {\bf 2} representation.
We expect that in an
extension of the SM this charge quantisation relation or some
generalisation of it will hold.

An obvious way of extending the SM is to extend the gauge group. The Standard
Model Group ($SMG$) is \cite{OrigOfSym,ORaif}
\begin{equation}
\label{SMGroup}
SMG \equiv S(U(2) \otimes U(3)) = U(1) \otimes SU(2) \otimes SU(3) /\hat{D}_{3}
\end{equation}
where the discrete group
\begin{equation}
\label{SMDiscrete}
\hat{D}_{3}\equiv \{(e^{i2\pi/6},-I_{2},e^{i2\pi/3}I_{3})^{n}:n\in
{\cal Z}_{6} \}
\end{equation}
ensures the above quantisation rule ($I_{N}$ is the identity of $SU(N)$).
We argue that
the most obvious extension is to add more groups to the sequence
$ U(1) \otimes SU(2) \otimes SU(3) $ and to use a different discrete group
so that the quantisation rule above is generalised to involve all the group
components. One of the groups we
consider is
\begin{equation}
\label{SMG235Group}
G_{5} \equiv U(1) \otimes SU(2) \otimes SU(3) \otimes SU(5) / \hat{D}_{5}
\end{equation}
where the discrete group $\hat{D}_{5}$ is defined as
\begin{equation}
\label{SMG235Discrete}
\hat{D}_{5} \equiv \{(e^{i2\pi/N_{5}},-I_{2},e^{i2\pi/3}I_{3},
e^{i2\pi m_{5}/5}I_{5}
)^n:n\in {\cal Z}_{N_{5}} \}
\end{equation}
where $N_{5}=2*3*5$ and $m_5$ is an integer which is not a multiple
of $5$.
%This is the simplest extension of the SM since we require that all '$N$'s
%(corresponding to the $SU(N)$ groups) are mutually prime.
This group gives a generalised quantisation rule
\begin{eqnarray}
\label{ChQuant235}
\frac{y}{2}+\frac{1}{2}\rm{``duality"}+\frac{1}{3}\rm{``triality"}+
\frac{m_{5}}{5}\rm{``quintality"} & \equiv & 0 \pmod 1
\end{eqnarray}
which is the simplest generalisation of the SM charge quantisation rule.
Further generalisations are obtained by extending the sequence
$U(1) \otimes SU(2) \otimes SU(3)$ with a set of $SU(N)$ factors, where the
`$N$'s are greater than 3 and mutually prime. The latter
condition ensures that the generalised quantisation rule shares the property
with the SM rule, eq.~(\ref{SMchqu}), that a given allowed value of
$\frac{y}{2}$ implies a unique combination of $N$-alities:
(duality, triality, \ldots, $N_i$-ality, \ldots)
\footnote{This corresponds to the global group, associated with
the generalised charge quantisation rule, having a connected centre
\cite{OrigOfSym}.}.

We will consider the fundamental scale to be the Planck mass
($M_{\rm Planck}$) and our models will
be a full description of physics without gravity below this scale. The
assumptions we make about our models essentially lead to the conclusion
that all new fermions with a
mass significantly below $M_{\rm Planck}$ must have a mass below the TeV scale
as explained in section~\ref{TechniColour}. Therefore our
models all describe low energy physics (below the TeV scale) and have a
desert up to the Planck scale where new physics will occur. We don't specify
any details about the Planck scale physics since it is largely irrelevant to
low energy physics.

We shall describe the gauge groups considered in this paper and the motivation
for choosing such groups in section~\ref{Groups}. We shall consider general
types of gauge groups and also give specific examples, concentrating on the
group $G_5$ defined above. When we also impose the condition
that all fermions are in fundamental
representations, as in the SM, we are limited to the models which we shall
consider in this paper. After choosing the gauge group we want to examine which
low mass fermions (low relative to the Planck scale) can exist in the model. We
must check that the model is then consistent, both theoretically and
experimentally.

The main theoretical constraint is that there are no anomalies as described in
section~\ref{Anoms}. This greatly limits the choice of fermions and their
weak hypercharges in our models. In appendix~\ref{SMGen} we show how the SM
generation can be derived using our assumptions about charge quantisation and
anomaly cancellation.

There is one important fact to keep in mind when proposing any extended model
which has extra non-Abelian gauge groups such as $SU(N)$. As we already know
from the SM, the $SU(3)$ group acts as a technicolour group \cite{TC} and
gives a contribution to the $W^{\pm}$ and $Z^{0}$ masses. In the SM this
contribution is very small but when confining groups with $N>3$ are considered
we must carefully
consider the effect this will have. Since we are not wanting the complications
of extended technicolour in order to generate quark and lepton masses,
we assume that there is a Higgs doublet and that the masses of the weak
gauge bosons are generated by a combination of the Higgs sector of the theory
and the technicolour effects of the gauge groups. This happens in exactly the
same way as in the SM where QCD gives a small contribution to the $W^{\pm}$
and $Z^0$ masses \cite{TC}.
% As discussed in section~\ref{Exp} this leads to a confinement scale
%for an $SU(N)$ group of less than 1TeV.

%By assuming that all fermions have a Yukawa coupling less than 10 at the
%electroweak scale, the upper limit for the mass of the fermions in our
%models is $\sim 1$ TeV.
For our models to be perturbatively valid, all Yukawa couplings at the
electroweak scale must be not much greater than 1. However, we will sometimes
take a somewhat higher mass threshold for all the new fermions when checking
to see if a model could be perturbatively valid up to the Planck scale.
For example, we can calculate the running gauge
coupling constants, assuming that all the new fermions can be included in the
renormalisation group equations (RGEs) at the TeV scale. Thus we can
check to see if any gauge coupling constant becomes infinite below the Planck
scale (i.e.\ if there are any Landau poles, especially for
the $U(1)$ coupling).
If the threshold was lower then the new fermions would effect the coupling
constants even more but this would only be a small effect.  Obviously we do
not want the coupling constants to become infinite or
the theory will be inconsistent. When we do this we find that there are few
self-consistent models allowed by our assumptions, in the sense that for any
particular gauge group only a few combinations of fermions which cancel
the anomalies do not cause the $U(1)$ gauge coupling to diverge.

%*******************
% The $SMG$ is given by $G_{3}$ so the
%minimal extension in this model is to consider the group
%\begin{equation}
%G_{5} \equiv U(1) \otimes SU(2) \otimes SU(3) \otimes SU(5) / D_{5}
%\end{equation}
%where the group $D_{5}$ ensures the generalised quantisation rule
%\begin{equation}
%\frac{y}{2} + \frac{1}{2}"duality" + \frac{1}{3}"triality" +
%\frac{m}{5}"quintality" \equiv 0\pmod{1}
%\end{equation}
%where $m$ is an integer not divisible by 5.
%*******************

We will show that in the model with gauge group $G_5$ we can add new
fermions with masses accessible
to present or planned future accelerators, in particular a fourth generation
of quarks without any new leptons.
%However the model itself makes
%no other predictions and explains no features of the SM\@.
Although the
model is consistent and can be tested experimentally in the near future,
it is not called for theoretically and does not resolve any of the outstanding
problems of the SM\@. Nevertheless it is the simplest alternative to the SM
which has the same characteristic properties as the SM itself.
%; in this sense
%it is the most `dull' extension of the SM\@.

In section~\ref{ThConstraints} we shall outline our requirements for a viable
model. We will discuss theoretical constraints such as anomaly
cancellation as well as aesthetic extrapolations from the SM,
including charge quantisation as already mentioned.

In section~\ref{ExpConstraints} we shall discuss the experimental constraints
which arise from the consistency of the SM with experiments. This includes the
experimental limits on the mass of the top quark and the masses of new,
undetected fermions.

In section~\ref{Simplifications} we will discuss the simplification of the
anomaly constraints when we assume that all fermions get a mass by the SM
Higgs mechanism.

In section~\ref{Only5sFails} we shall show the difficulty of constructing
a model where all the new fermions are in 5-plet or anti-5-plet
representations of $SU(5)$. We shall show that such a solution is not
possible within the context of our model.

In section~\ref{SU2*SUN} we will see how the difficulties of section
\ref{Only5sFails} can be overcome by also adding fermions which are $SU(5)$
singlets; in particular a fourth generation of quarks but no fourth
generation of leptons. We will also show how such a solution can be
formulated in a more general gauge group.

In section~\ref{Conclusion} we shall discuss the overall merits of such
a model and how easily it could be tested experimentally.

%********************
%
%
\section{Discussion of Formalism of Models
and Theoretical Constraints}
\label{ThConstraints}

First we shall discuss which models we will be considering as viable extensions
of the SM and then
we shall discuss in detail the requirements for a potentially successful
extension of the SM\@. We shall use some of these constraints when constructing
models and the rest to check the consistency of our models.

\subsection{Extrapolations From the SM}
\label{SMExtrapolations}

In this section we discuss aesthetic extrapolations from the SM\@.
These are features of
the SM which have no obvious explanation but in some way can be used to specify
the model almost uniquely. We try to pick out these features and carry
them over
to or generalise them in our extended model. This is a method of selecting a
particular model and our view is that this is the most logical method although
the features chosen may of course be subject to personal prejudice.

\subsubsection{Extending the Gauge Group and Charge Quantisation}
\label{Groups}
\label{ChQuant}
%
%**********************
%\input{groups}
As stated in section~\ref{Intro}, an obvious way of extending the SM is to
extend the gauge group. The $SMG$ is;
\begin{equation}
SMG \equiv U(1) \otimes SU(2) \otimes SU(3) / \hat{D}_{3}
\end{equation}
where the discrete group
\begin{equation}
\hat{D}_{3}\equiv \{(e^{i2\pi/6},-I_{2},e^{i2\pi/3}I_{3})^{n}:n\in {\cal Z}_{6}
\}
\end{equation}
ensures the quantisation rule, eq.~(\ref{SMchqu}). We believe that the most
obvious extension is to add more special unitary groups to the sequence
$ U(1) \otimes SU(2) \otimes SU(3) $ and to use a different discrete group
so that the quantisation rule above is generalised. In \cite{OrigOfSym} it
is argued that the group should be of the form
\begin{equation}
G_{p} \equiv U(1) \otimes SU(2) \otimes SU(3) \otimes SU(5) \otimes
\cdots \otimes SU(p) / \hat{D}_{p}
\end{equation}
where the product is over all $ SU(q) $ where $q$ is a prime number less
than or equal to the prime number $p$. The discrete group $\hat{D}_{p}$ is
defined as
\begin{equation}
\hat{D}_{p} \equiv \{(e^{i2\pi/N_{p}},-I_{2},e^{i2\pi/3}I_{3},
e^{i2\pi m_{5}/5}I_{5},\ldots,
e^{i2\pi m_{p}/p}I_{p})^n:n\in {\cal Z}_{N_{p}} \}
\end{equation}
where $N_{p}=2*3*5* \cdots *p$ and $m_N$ is an integer which is not a multiple
of $N$. In fact we can obviously choose $0 \le m_N \le N-1$ since $m_N$ is
really only defined modulo N. We also have the freedom to choose that there
are, for example, at least as many $SU(2)$ doublets which are ${\bf N}$
representations of $SU(N)$ as ${\bf \overline{N}}$ representations since we can
conjugate $SU(N)$ and set $m_N \rightarrow -m_N \pmod N$. We will use this fact
later to eliminate duplicate solutions where all N-plets and anti-N-plets have
been interchanged. This also allows us to fix $m_3=1$.

This group gives a generalised quantisation rule
\begin{eqnarray}
\frac{y}{2}+\frac{1}{2}\rm{``duality"}+\frac{1}{3}\rm{``triality"}+ & &
\nonumber \\
\frac{m_{5}}{5}\rm{``quintality"}+\cdots+\frac{m_{p}}{p}\rm{``p-ality"} &
\equiv & 0 \pmod 1
\end{eqnarray}

We will also consider the more general groups defined as
\begin{equation}
\label{SMG2NNGroup}
SMG_{2N_1N_2 \ldots N_k} \equiv U(1) \otimes SU(2) \otimes SU(N_1) \otimes
\cdots
\otimes SU(N_k) / D_{2N_{1} \ldots N_{k}}
\end{equation}
where
\begin{equation}
\label{SMG2NNDiscrete}
D_{2N_{1} \ldots N_{k}} \equiv \{(e^{i2\pi/\hat{N}},-I_{2},e^{i2\pi
m_{N_1}/N_1}I_{N_1},
\ldots ,e^{i2\pi m_{N_k}/N_k}I_{N_k})^m
:m \in {\cal Z}_{\hat{N}} \}
\end{equation}
Here $\hat{N}=2*N_1* \cdots *N_k$ and the $N_i$ are odd and mutually prime
(we can obviously assume they are arranged in ascending order).
So the quantisation rule is
\begin{equation}
\label{GenChQuant}
\frac{y}{2}+\frac{1}{2}d+\frac{m_{\scriptscriptstyle{N_1}}}{N_1}n_1+ \cdots
+\frac{m_{\scriptscriptstyle{N_k}}}{N_k}n_k \equiv 0 \pmod 1
\end{equation}
where we have defined $d$ to be the duality and $n_i$ to be the $N_i$-ality of
a representation. The groups $SMG_{23N}$ are the minimal extensions of the
$SMG$ ($\equiv SMG_{23}$) which are inspired by the $SMG$, in the sense that
each is also a cross product of $U(1)$ and a set of distinct special unitary
groups with a charge
quantisation rule involving all the direct factors and contains
the $SMG$ as a subgroup. The property of the $SMG$ that the value of
$\frac{y}{2}$ determines both the duality and triality extrapolates to the
principle that $\frac{y}{2}$ should also fix the $N$-ality, but then it is
needed that 2, 3 and $N$ are mutually prime.

It has been suggested that a defining property of the SMG is that it has few
outer automorphisms relative to the rank of the group \cite{MostSkew,skewchi}.
This can be described by saying that it is very skew.
The intermingling of the various simple groups $SU(2)$, $SU(N_1)$,
\ldots $SU(N_k)$ implied by the charge quantisation rule,
eq.~(\ref{GenChQuant}), helps
to suppress the number of outer automorphisms and ``generalised
automorphisms''.
Thus a group like $SMG_{2N_1N_2 \ldots N_k}$ would indeed be more skew than
groups without such intermingling.
%If we accept this principle, which is suggested by random dynamics
%\cite{OrigOfSym}, then the groups $SMG_{2N_1N_2 \ldots N_k}$
%are naturally suggested as alternatives to the $SMG$. In particular, the
%requirement that all the $N_i$ be mutually prime and the definition of the
%discrete group $D_{2 N_1 N_2 \ldots N_k}$ follow from this principle.
Alternatively we can derive eq.~(\ref{GenChQuant}) directly as a natural
generalisation of the SM charge quantisation rule, eq.~(\ref{SMchqu}).
%In particular this would
%mean that all the $N_i$ were distinct odd primes. The discrete group would
%%also
%be defined although we can also derive that directly as an extension of the SM
%charge quantisation rule.

Of course it is possible that the apparent charge quantisation rule in the SM
is simply due to chance; i.e.\ the fermions in the SM just happen to obey that
particular rule. However we believe that the quantisation rule is a fundamental
feature of the SM; so we argue that
it is very difficult to see how
there cannot be a generalisation of this rule in an extended model, while
still retaining the general features of the SM\@. In fact the form of the
generalised quantisation rule is suggested from the SM and there seems to be
little choice
in selecting the rule since the SM rule appears to be the one which involves
all the direct factors equivalently. In fact the choice of the most complicated
charge quantisation rule in some way defines the $SMG$. This is why we have
divided out the discrete
groups $\hat{D}_{p}$ and $D_{2N_{1} \ldots N_{k}}$.

%**********************

\subsubsection{Small Representations}
\label{SmallReps}

In the SM, for each $SU(N)$ group, the fermion representations are
either N-plet
(${\bf N}$), anti N-plet (${\bf \overline{N}}$) or singlet ({\bf 1}).
This can be described by
saying that all the fermions lie in fundamental representations of each $SU(N)$
group to which they couple. We pick this as a feature of the SM
which we shall extend to our models.
We note here that this is in contrast to some other attempts
to extend the SM\@.
For example in SUSY there are fermions in other representations
(e.g.\ gauginos in adjoint representations).
Fundamental representations are also suggested in \cite{Why3+1} since these
make the Weyl equation most stable when considering random dynamics
\footnote{In fact, from this point of view, each representation
of the full gauge group should only be non-singlet with respect to one
non-Abelian factor. This is not true for the left handed quarks but is true
for all other fermions in the SM\@. However the left handed quarks are required
in order that there are no gauge anomalies. So we can consider that the
Weyl equation is as stable as possible if we only have small representations.}.

Another feature is that the weak hypercharge is in some way minimised in
the SM, subject of course to the constraints of anomaly cancellation and charge
quantisation, as shown in appendix~\ref{SMGen}. So in our extended model we
will choose hypercharge values close to zero when this is possible.
More precisely, we choose to minimise the sum of weak hypercharges
squared over all fermions.
This will also
minimise the running of the $U(1)$ gauge coupling constant
%effect the $U(1)$ running coupling constant the least
and so give each model the best chance of being consistent up to the
Planck scale, which we require as stated in section~\ref{Desert}.

\subsubsection{Higher Energies - Desert Hypothesis}
\label{Desert}

The SM has been tested at energies up to a few hundred GeV. There have been
many theories proposed which would be valid at energy scales ranging from 1 TeV
up to the Planck scale around $10^{16}$ TeV. Many of these
theories have a large
range of energy where no new physics occurs. One example is GUTs where there
is typically no new physics from the SM energy scale up to the grand
unification scale around $10^{13}$ TeV. An alternative is that there is no new
physics until the Planck scale where we can be almost certain that quantum
gravity will have a significant effect. We shall adopt this view for our
extended
models. This means that once we have set the mass scale for the fermions in the
extended model, we can calculate the running coupling constants and check to
see if there is a Landau pole below the Planck scale, i.e.\ whether the $U(1)$
gauge coupling becomes infinite below the Planck scale.
If there is a Landau pole
then we will conclude that such a model is not consistent.

\subsection{Fermion Representations and Alternative Groups}
\label{AltGroups}

%*******************
%\input{altgroup}
In this section we shall describe some alternative extensions of the SM. We
will consider groups similar to those we are examining in this paper in the
sense that they contain the $SMG$ and additional special unitary group factors.
This obviously does not include models which unify the individual components
of the $SMG$ or models which involve SUSY. There have been many such models
and the additional symmetries are usually used to explain coupling constant
unification, the number of families in the SM or the fermion mass hierarchy in
a fairly natural way.

In the models described in section \ref{Groups} the SM fermions cannot couple
to any new gauge fields because
of the charge quantisation rule. This is due to the fact that all values of
$\frac{y}{2}$ in the SM are multiples of $\frac{1}{6}$ and so the charge
quantisation rule, eq.~(\ref{GenChQuant}), forces the SM fermions to be
singlets of all $SU(N)$ groups where $N>3$ are distinct primes.

However the situation is more complicated if we allow more than one $SU(N)$
gauge group for any particular $N$. Where we have $N=2$ or 3 there are two
distinct cases. In the first case the SM group $SU(N)$ is an
invariant subgroup of the extended group. We then call the extra $SU(N)$ groups
a horizontal symmetry. In the other case the $SU(N)$ group in the $SMG$ is not
an invariant subgroup and is generally a diagonal subgroup of the extended
group.

\subsubsection{Invariant Subgroup Case: Horizontal Symmetries}
If we have one more $SU(2)$ or $SU(3)$ group then we can have a
horizontal symmetry (a non-abelian symmetry which places fermions from
different generations in the same multiplet). The idea of a gauged horizontal
symmetry is not new and has been used to try
and explain the mass hierarchy of the SM fermions \cite{HSym}. However, an
$SU(N)$ group with $N>3$
is not a possible horizontal symmetry without introducing many more fermions
because there are only 3 generations of SM fermions and the smallest
non-trivial representation of $SU(N)$ is the N-plet.
For example if $N=5$ we would have an $SU(5)$ horizontal symmetry and so
we would need at least 5 generations of SM fermions.

If the horizontal symmetry gauge group is $SU(3)_H$ then we must place fermions
from different generations in the same triplet (or anti-triplet). It turns out
that the only way to do this, avoiding anomalies (see section~\ref{Anoms}) and
not introducing any new fermions, is to put all fermions in the same (or
conjugate) representation of $SU(3)_H$ as they are in the colour group
$SU(3)_C$ of the SM; so that all three generations of left handed quarks are
put in a triplet (or anti-triplet) of $SU(3)_H$ etc. However, the SM fermions
would not then obey the charge quantisation rule which might be expected,
similar to eq.~(\ref{GenChQuant});
\begin{equation}
\frac{y}{2}+\frac{1}{2}d+\frac{1}{3}t_C+\frac{1}{3}t_H \equiv 0 \pmod{1}
\end{equation}

If the horizontal symmetry group is $SU(2)_H$ then we can make some or all SM
fermions triplets of $SU(2)_H$ but this is not the smallest representation and
so we do not favour this as explained in section~\ref{SmallReps}. We could
place some fermions in doublets of $SU(2)_H$. This could be done, without
introducing any anomalies, by placing two generations of quarks in the same
doublet or taking two generations and placing the fermions in the same
representation of $SU(2)_H$ as they are in $SU(2)_{L}$. Different doublets
could connect fermions from a different pair of generations. For example left
handed quarks from the first and second generations
could be in the same doublet, right handed `up' quarks from the first and third
generations could be in the same doublet and right handed `down' quarks from
the second and third generations could be in the same doublet. This would not
give any anomalies though it is difficult to see how this could be used to
explain the fermion masses.
The main problem is that fermions in different
generations with very different masses are put in the same multiplet. This
means that the fermions would naturally get the same mass. It is difficult to
break the symmetry in such a way that the masses of all the different fermions
are split by realistic amounts \cite{HSym}.

We do not consider these
possibilities in this paper because triplets of $SU(2)$ are not fundamental
representations and the other possibilities, with gauge group $SU(2)_H$ or
$SU(3)_H$, mean that the fermions could not obey
the extended charge quantisation rule. Of course models involving horizontal
symmetries do not enforce such charge quantisation rules.

\subsubsection{Non-invariant Subgroup Case: $SMG$ as Diagonal Subgroup}

In the case where, for example, the $SU(3)_C$ subgroup of the $SMG$ is not an
invariant subgroup of the full gauge group, the only possibility is that it is
a diagonal (or anti-diagonal) subgroup of $SU(3)^n$. In this type of model
different generations can couple to different $SU(2)$ and $SU(3)$ gauge groups
in the full gauge group. There would then be symmetry breaking to produce the
$SMG$, in such a way that $SU(3)_C$ could be said to be a diagonal subgroup of
all the $SU(3)$ groups in the full group which exists at energies higher than
the symmetry breaking scale. In other words, $SU(3)_C$ is then the subgroup in
which all the $SU(3)$ groups undergo the same transformations. In this way it
is trivial to cancel all the anomalies, since each generation of quarks and
leptons cancel all anomalies separately and couple to a $U(1) \otimes SU(2)
\otimes SU(3)$ subgroup of the full group in the same way as they couple to the
$SMG$. This is in contrast to the invariant subgroup case, where the SM
fermions
had to couple to the $SMG$ and also to other subgroups of the full gauge group.
Also in the diagonal case, the dimension of each representation is the same as
in the SM, whereas, in the invariant subgroup case, the dimensions were larger
since different SM representations were combined under the horizontal symmetry.

This type of model has been proposed \cite{Diag} as an alternative to
horizontal symmetries or grand unification. Examples include topcolour models
\cite{TopColour} and the anti-grand unification model \cite{AntiGUT}, where the
group $SMG^3 \equiv SMG \otimes SMG \otimes SMG$ has been used to successfully
predict the values of the gauge coupling constants.
The anti-grand unification model has also
been analysed as a model to explain the hierarchy of
SM fermion masses \cite{SMG3}. Here the extended model with gauge group
$SMG^3 \otimes U(1)_f$ has been fairly successful at reproducing the observed
fermion masses in an order of magnitude approximation (reproducing all SM
fermion masses within a factor of 2 or 3). The extra $U(1)_f$ gauge symmetry is
called a flavour symmetry and is
required to produce the observed mass differences within the second and third
generations, e.g. $m_b \ll m_t$.

We note that the fermions in some of these models obey extended charge
quantisation rules which we would expect. For example the fermions in the
$SMG^3$ model obey the charge quantisation rules;
\begin{equation}
\frac{y_i}{2}+\frac{1}{2}d_i+\frac{1}{3}t_i \equiv 0 \pmod 1
\end{equation}
where the three copies of the $SMG$ are labelled by $i=$ 1, 2 and 3. With three
separate charge quantisation rules, this is not truly a straightforward
extrapolation of the SM charge quantisation rule. However it is similar in
the sense that these rules are required to produce the group $SMG^3$ which has
as large $\chi$ as the group $SMG$ itself.
\footnote{The quantity $\chi$ is defined in \cite{skewchi} for any group $G$ as
$\chi(G)=\ln(q(G))/r(G)$ where $r(G)$ is the rank of the group $G$
(really the number of $U(1)$ factors in the maximal abelian subgroup). Further,
$q(G)$ is defined as the order of the factor group, obtained by dividing the
group of all abelian charge combinations $(y_1,y_2, \ldots ,y_r)$ allowed for
any representations of the group $G$, by the group of those abelian charge
combinations allowed for representations trivial under the semi-simple part of
the group $G$.}
The quantity $\chi$ measures how strongly
intermingled the $U(1)$ subgroups are with the semi-simple part via the
dividing discrete groups (i.e. equivalently via
the quantisation rule(s)). It happens that groups of the form $SMG^n$
have the largest possible value of this measure $\chi$= $\ln(6)/4$.
The charge quantisation rules are chosen to maximise
$\chi$ for the group $SMG^3 \otimes U(1)_f$ among all those with the same
algebra although this group does not have as large a value of $\chi$ as the
$SMG$. In fact $\chi=\ln(6^3)/13=\frac{12}{13}\ln(6)/4$ for the
group $SMG^3 \otimes U(1)_f$.

However, the symmetry breaking scale of the group $SMG^3$ is taken to be just
below the Planck scale in the anti-grand unification model and in this paper we
wish to study the possibilities of new physics at much lower energies; energies
of the same order of magnitude as the electroweak scale rather than the Planck
scale. This is still possible in such a model but it then loses its ability to
%reproduce the fermion masses and
predict the gauge coupling constants. Topcolour models do introduce new
dynamics at the TeV scale but in this paper we shall not consider such models.

%*******************

\subsection{Anomalies}
\label{Anoms}

\subsubsection{Gauge Anomalies}
\label{GaugeAnoms}

In any chiral gauge theory, gauge anomalies can arise. These anomalies lead to
an inconsistent theory and so they must not be present in a good theory. Each
fermion representation makes its contribution to each type of anomaly. We
say that there is an anomaly present if the total contribution to an
anomaly from all the fermion representations is non-zero.

As explained in section~\ref{Groups}, the models considered in this paper have
gauge groups of the general form
\begin{equation}
U(1) \otimes \prod_{i} SU(N_{i}) /D \nonumber
\end{equation}
The discrete group $D$ leads to charge quantisation. We assume all fermions
to be in ${\bf N}$, ${\bf \overline{N}}$ or singlet ({\bf 1})
representations of each $SU(N)$, as
discussed in section~\ref{SmallReps}.
We define n to be the N-ality of a representation
($n=1$ (-1) for representation ${\bf N}$ (${\bf \overline{N}}$)
and $n=0$ for singlet
representation). We can also define the size, $S$, of each representation as
the dimension of the representation (e.g.\ in the SM, $S=6$ for the $(2,3)$
representation of $SU(2) \otimes
SU(3)$ which is equivalent to the fact that there are 6 left-handed quarks
in each generation).

For gauge anomalies we sum the contribution for all left-handed fermions and
subtract the sum over all right-handed fermions. This is equivalent to summing
over left-handed fermions and left-handed anti-fermions.
We have now introduced all
the necessary notation to write down general equations for all types of gauge
anomalies.
%
%****************
\begin{figure}
\epsfxsize=\textwidth
\epsffile[0 500 596 843]{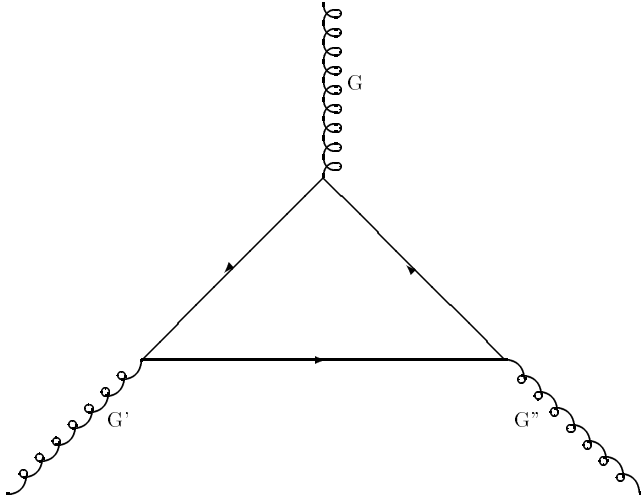}
\caption{For the theory to be anomaly-free, the amplitude of this Feynman
diagram must be zero for all choices of
external gauge bosons after summing over all possible fermions in the internal
loop (triangle).}
\label{AnomDiag}
\end{figure}
%****************

The requirement that there are no anomalies present in a theory is analogous
to the triangle Feynman diagram in Fig.~\ref{AnomDiag} with a fermion loop
and three external gauge
bosons (labelled by $G$, $G'$ and $G''$) having zero amplitude for all possible
choices
of gauge bosons $G$, $G'$ and $G''$. The contribution from
each fermion representation
is calculated by making particular choices for the fermions in the internal
loop.
These contributions must then sum to give zero amplitude if there is to be
no anomaly.

If each of $G$, $G'$ and $G''$ is an $SU(N)$ gauge boson
where $N \ge 3$ then each
representation gives a relative contribution of $Sn^{3}=Sn$ (since $n=-1$, 0
or 1 in our models). The total contribution is therefore
$\sum_{i}S_{i}n_{i}$ where $i$
labels each left-handed fermion (and anti-fermion) representation.
We label this type of anomaly $[SU(N)]^{3}$ and require
\begin{equation}
\sum_{i}S_{i}n_{i}=0
\end{equation}

Another type of anomaly corresponds to the diagram with one $U(1)$ gauge boson
and two $SU(N)$ gauge bosons where $N \ge 2$, labelled as $[SU(N)]^{2}U(1)$.
Each representation gives a relative contribution $Sn^{2}y$. Therefore we
require
\begin{equation}
\sum_{i}S_{i}(n_{i})^{2}y_{i}=0
\end{equation}

The final type of gauge anomaly corresponds to the diagram with all the
gauge bosons $G$, $G'$ and $G''$ being $U(1)$ gauge bosons. This is labelled as
$[U(1)]^{3}$ and each representation gives a relative contribution $Sy^{3}$.
Therefore we require
\begin{equation}
\sum_{i}S_{i}y_{i}^{3}=0
\end{equation}

\subsubsection{Other Anomalies}

There is also a mixed gravitational and gauge anomaly \cite{MixedAnom} which
corresponds to one $U(1)$ gauge boson and two gravitons. We will label this
as $[{\rm Grav}]^2U(1)$. Each representation gives a relative
contribution $Sy$ and so this leads to the constraint
\begin{equation}
\sum_{i}S_{i}y_{i}=0
\end{equation}

Another possible anomaly is the Witten discrete $SU(2)$ anomaly
\cite{WittenAnom}. This states
that if the number of left-handed $SU(2)$ doublets is odd then the theory
is inconsistent. As we shall see later this anomaly does not give us any
problems.

\section{Experimental Constraints}
%\section{Constraints on the Fermion Content of Our Models}
\label{ExpConstraints}

In this section we shall discuss the constraints on our models which are due
to experimental evidence. In particular we are concerned with the possibilities
for the existence of more fermions and what restrictions can be imposed both
directly and indirectly on their mass. Some difficulty arises since fermions
may be confined and so not directly observable. This means that direct
experimental
restrictions will refer to the mass of particles which are combinations of
these fermions, like hadrons in the case for quarks.
%, since we don't want to make any assumptions about decay modes.

\subsection{Direct Experimental Constraints on Fermion Masses}

First we shall discuss the constraints on fermion masses due to the fact that
so far no non-SM fermions have been observed. We shall show that this rules out
any extra massless fermions and then give current limits on the masses of
different type of new fermions.

%****************
%\input{smgexp}
\subsubsection{Massless Fermions}
\label{masslessfermions}

Only three massless fermions have been observed and they are the three massless
neutrinos described in the SM (even if the neutrinos do have a small mass
we know that there are only 3 with a mass less than $\frac{1}{2}M_Z$). Any
other massless fermions, which had any significant
coupling to the SM fermions or gauge bosons, would have been observed if they
were not confined. When we assume that fermions belong only to fundamental
and singlet representations (as postulated in section~\ref{SmallReps}), the
charge quantisation rule in our models ensures
that the only possible fermions which would not be electrically charged
%(and couple to the electroweak gauge bosons)
would be neutrinos. A left-handed neutrino without a right-handed neutrino
would be massless as in the SM. We already know that there are only three such
neutrinos and so we cannot consider this as a possibility for new fermions. A
right-handed neutrino would be completely decoupled from the gauge group and
so it could get a gauge invariant Majorana mass. So we
would expect that it would have a mass $\sim M_{\rm Planck}$
and so it is excluded
as a low mass fermion in our models. Therefore any new massless fermions in our
models must be electrically charged and so must also be confined by a new
interaction well above the QCD scale, on phenomenological grounds.
% Also, even if we allowed
%a right-handed neutrino it could not be used to give the left-handed neutrino
%a significant mass by the usual SM Higgs mechanism. For this reason we cannot
%allow any more generations of SM leptons since we cannot give an explanation
%for a neutrino mass larger than $\frac{1}{2}M_Z$.

If there is a confined gauge group then we assume that fermion condensates will
be formed as in QCD. If a fermion doesn't have a chiral partner with respect to
some confined group $H$, the condensates formed will break the group $H$.
So if we assume that there is no spontaneous gauge symmetry breaking, other
than that of the electroweak
symmetry group, no fermions can be chiral w.r.t.\ $G$ where the full gauge
group is $U(1) \otimes SU(2) \otimes G/D$ (where $D$ is some discrete group).
In our models the extra $SU(N)$ gauge
groups are all confining (with negative beta functions), so that $G \equiv H$.

If the left- and right-handed fermions occur with the same representations of
the full gauge group $U(1) \otimes SU(2) \otimes G/D$, then the fermions can
form a Dirac mass term in the Lagrangian. So they would be expected to get a
mass comparable to the fundamental scale, which we take to be the Planck mass
in our models. Such fermions would not contribute to any anomalies and would
not be observable because of their high mass. We shall therefore ignore them in
our models. If a fermion cannot form such a fundamental Dirac
(or Majorana) mass
term then we say it is mass protected, since it would be fundamentally massless
and could only get a mass indirectly through some interaction such as the Higgs
mechanism. All the fermions considered in our models are mass
protected by the electroweak interactions.

We conclude that all new fermions in our models must get their mass from the
Higgs mechanism. Furthermore, they must couple to the usual SM Higgs particle
in the same way as the
SM fermions. In other words, the fermion condensates must have the same
quantum numbers as the SM Higgs boson; otherwise their contributions to the
$W^{\pm}$ and $Z^0$ masses, via the usual technicolour \cite{TC} mechanism,
would be analogous to those from the vacuum expectation values of Higgs
particles with non-standard weak isospin and hypercharges.
%the technicolour effect described in
%section~\ref{TechniColour} would effect the $W^{\pm}$ and $Z^{0}$ masses in a
%different way from the predicted effects of the standard Higgs
%electroweak symmetry breaking.
This would lead to a significant deviation of the $\rho$ parameter
($\rho \equiv \frac{M_W^2}{M_Z^2 cos^2\theta_W}$) from unity \cite{rho=1} in
contradiction with precision electroweak data.

%We know from accelerator experiments with great precision and also less
%precisely from cosmological arguments that there are just three massless
%neutrinos (certainly with mass less than 31MeV). In the SM every generation
%has a massless neutrino because there are no right-handed neutrinos and so
%we can conclude that there are precisely three generations.
%
%The experimental evidence obviously forbids any other massless fermions which
%are not confined (i.e.\ any other massless leptons). However it may still be
%possible to have massless fermions if they are confined. In this case we must
%try and estimate the mass of the particles which would be seen. Since neutral
%particles are much more difficult to observe we shall consider the charged
%`hadrons' which would be formed. If the fermion was neutral then this would
%obviously not be possible but in this paper all our models have a charge
%quantisation rule which does not allow any neutral fermions other than
%leptons.
%
%It is not entirely clear whether a massless fermion would give rise to a
%massless `hadron' or not. If it did then it would certainly have been detected
%already but if for some reason the `hadrons' had masses then it is possible
%that they may not have been detected.

\subsubsection{Massive Fermions}
\label{expmass}

In the SM there are two different types of fermions, quarks and leptons, which
differ by the fact that quarks couple to the $SU(3)$ gauge fields and so are
confined, whereas leptons have no direct coupling to the $SU(3)$ gauge fields
and are not confined. There are experimental limits on the masses of any quarks
and leptons which have not yet been observed. If there are any more leptons
then they must have a mass greater than 45 GeV \cite{PDG}. We shall
assume that there are
no more leptons, since even the neutrino would have to get a mass larger than
this and it is
difficult to see how a neutrino could naturally be given a mass greater than 45
GeV but still much lower than the fundamental scale (which is the Planck scale
in our models). This is because a right-handed neutrino, as already discussed
in section~\ref{masslessfermions}, would naturally get a Majorana mass and so
the see-saw mechanism \cite{See-Saw} would leave the left-handed neutrino with
a very small mass. For this reason we cannot allow any more generations of SM
leptons. However the limits on the quark masses are dependent on the type of
quark and its decay modes.

%It has recently been reported that the top quark has been observed with a
%mass of 174GeV. However this
%is only a preliminary finding and until it is confirmed the experimental limit
%is $m_t \ge 130GeV$. We also note that this limit would obviously apply to a
%fourth generation `up-type' quark which we would label $t'$
%but the experimental
%limit on the corresponding $b'$ quark is not as high, $m_{b'} \ge 85GeV$.

The top quark has recently been observed by the CDF
%\cite{CDFtop}
and D0
%\cite{D0top}
collaborations \cite{CDFD0top}. The mass is in the range 150-220 GeV.
% The
%cross-section is higher than the theoretical expectations though still
%consistent considering the size of the errors. However it is also consistent
%with a fourth generation up quark, t', with $m_{t'} \sim m_t$.
%However this limit wouldn't apply to a 4th generation up-type quark, t', since
%the search for the top looked for b jets and these wouldn't be observed
%from the decay of a t' to a heavy b'. The current limits on a t' quark are
%therefore not clear but previous searches for the top suggest a limit of
%$m_{t'}>130$ GeV.
%There is a limit on a fourth generation down quark, b', though.
%From \cite{PDG}
%$m_{b'} \ge 85$ GeV.
For the purpose of this paper we take the limit on possible fourth generation
quarks, $t'$ and $b'$, to be
\begin{displaymath}
M_{t'}, M_{b'} > 130 \; \rm{GeV}
\end{displaymath}
from the dilepton analyses of the CDF
%\cite{CDF4gen}
and D0
%\cite{D04gen}
groups \cite{CDFD04gen}
(less restrictive limits apply if other decay modes are dominant).
Note that experimental limits are taken to apply to the pole masses.

%It is possible that fermions which were not singlets of additional
%gauge groups could have a mass lower than this since they would be
%confined and
%the mass of the `hadrons' would depend on the scale of confinement which could
%be much higher than the QCD scale. However we are only really concerned
%with providing an upper limit on such fermions since this can be used to check
%that if the model has a Landau pole below the Planck scale. If the model was
%consistent then we could use the RGEs to provide better limits
%by examining the
%infra-red fixed points of the Yukawa couplings.
%The experimental lower bounds can be used to constrain our models since we can
%make theoretical predictions on the upper limits of any fermions once we have
%specified a particular model.
%To do this we use the
%RGEs to examine how the Yukawa couplings evolve when we run the
%equations from the Planck scale down to the electro-weak scale. In the SM
%this provides an upper limit on $m_t$ and $m_H$, the mass of the Higgs scalar.
The above experimental limits do not apply to new fermions which are not
singlets of the additional $SU(N)$ gauge groups. These fermions would be more
difficult to detect experimentally and would anyway be confined inside
`hadrons' with a confinement scale (generically at the electroweak scale) much
higher than the QCD scale. We require our models to remain perturbative in the
desert from the TeV scale to the Planck scale. So we can use the RGEs to
examine how the Yukawa couplings evolve from the Planck scale down to the
electroweak scale. In particular we study the infra-red quasi-fixed-point
structure of the renormalisation group equations (RGEs).
In the SM the fixed point values provide upper limits on
the mass of the top quark, $M_t$, and the Higgs scalar, $M_H$.
Similarly in extended models we get upper limits on the masses of the heaviest
fermions, though the precise values depend on the relative masses of these
fermions and also the unknown gauge coupling strength, $g_N$, of the $SU(N)$
groups to which the fermions couple. Also we must be careful to point out that
the RGEs describe the running of the Yukawa couplings and, as we discuss in
section~\ref{TechniColour}, the actual masses will be less than naively
expected,
due to the technicolour-like contribution from $SU(N)$ to the electroweak
vacuum expectation value (VEV), $v=246 \rm{\;GeV}$. As we shall see, this will
enable us to quite accurately predict the masses of some of the fermions we
introduce in our model in section~\ref{SU2*SUN},
since we have theoretical upper
limits and experimental lower limits.

%****************

%*************
%\input{techni}
\subsection{Technicolour Contributions}
\label{TechniColour}

Technicolour theories \cite{TC} have been proposed as an alternative to the
Higgs
mechanism to provide a mass for the weak gauge bosons. This is based on the
fact that QCD would provide a (very small) mass for these bosons without
any Higgs scalars. Similarly any other confining $SU(N)$ gauge groups, with
fermions which are non-trivial under $U(1) \otimes SU(2)$, are expected to form
fermion condensates which would contribute
to the $W^{\pm}$ and $Z^0$ masses. In our models the charge quantisation rule
ensures that all fermions are non-trivial under $U(1)$. Thus all $SU(N)$
groups in our models, which are coupled to fermions, will contribute to the
weak
boson masses.

We stress that we are not proposing a technicolour model as such, but simply
taking into account the unavoidable effect that adding an $SU(N)$ group has.
We are assuming that the Higgs sector of our models is the same as in the SM,
i.e.\ one Higgs doublet, and that the fermion condensates have the same quantum
numbers as the Higgs doublet. Then the VEV due to the Higgs field,
$<\phi_{WS}>$, is related to the total VEV, $v$, and the contribution from
$SU(N)$ due to fermion condensates, $F_{\pi_N}$, by the relation:
\begin{equation}
<\phi_{WS}>^2 + F_{\pi_N}^2 = v^2 = (246 \; \rm{GeV})^2
\end{equation}
which is exactly the same as in
technicolour models with a scalar \cite{TechniHiggs}.

The fermion running masses, $m_f$, are related to the Higgs field VEV in the
usual way:
\begin{equation}
m_f = \frac{y_f}{\sqrt{2}}<\phi_{WS}>
\end{equation}
where $y_f$ is the Yukawa coupling constant for the fermion f ($y$ is used for
both Yukawa coupling and weak hypercharge but it should be obvious from the
context which is being referred to). For quarks, the running mass is related to
the pole
mass, $M_f$, by
\begin{equation}
\label{QuarkPole}
M_f=\left(1+\frac{4\alpha_S(M_f)}{3\pi}\right)m_f(M_f)
\end{equation}
where $\alpha_S(M_f)$ is the QCD fine structure constant at the pole mass.
For quarks with a mass of order $M_Z$ we can approximate
$\alpha_S(M_f) \approx \alpha_S(M_Z)$ to give the approximate formula:
\begin{equation}
M_f \approx 1.05m_f(M_f)
\end{equation}
This means that the pole mass of a heavy quark will be about 5\% higher than
the
running mass. However, we will use eq.~(\ref{QuarkPole}) when calculating the
pole masses of the quarks.

Using the Yukawa coupling infrared quasi-fixed point value as an upper bound,
we must avoid any significant suppression of the top quark and possible fourth
generation quark masses due to the reduction of $<\phi_{WS}>$ below its SM
value. We usually imagine taking
\begin{equation}
F_{\pi_N} \quad \leq 75 \; \rm{GeV}
\end{equation}
and thus
\begin{equation}
<\phi_{WS}> \qquad > \qquad 234 \; \rm{GeV}
\end{equation}
In fact we shall quote limits on fermion pole masses based on taking
\begin{equation}
<\phi_{WS}> \quad = \quad 234 \; \rm{GeV}
\end{equation}
This gives the following relation for the pole mass of quark $f$:
\begin{equation}
\label{PoleYuk}
M_f = \left( 1+\frac{4\alpha_S(M_f)}{3\pi} \right) \frac{y_f(M_f)}{\sqrt 2}
<\phi_{WS}>
\end{equation}
In the approximation $M_f \approx M_Z$ we get:
\begin{equation}
M_f \approx 174 y_f(M_f) \; {\rm GeV}
\end{equation}

Upper limits for fermion masses are obtained by using quasi-fixed-point values
for the
Yukawa coupling constants, $y_f$, as determined from the RGEs in viable models
with a desert above the TeV scale. These infra-red fixed point Yukawa couplings
are of order unity. However for the purposes of investigating the behaviour of
the gauge coupling constants, and especially to demonstrate  that the $U(1)$
coupling constant develops a Landau pole in our model without new fermions
(section~\ref{Only5sFails}), we take a more generous single threshold
of ten times the electroweak scale $\sim 1.7$ TeV for all new fermions in
that model. For our discussion in section~\ref{SU2*SUN} of the model with a
fourth generation of quarks we take the more stringent lower threshold value of
$M_Z$, in order to demonstrate the absence of Landau poles in this case.

\subsection{Precision Electroweak Data}
\label{Epsilons}

Measurements of electroweak interactions are now accurate enough to be
sensitive to loop corrections to propagators and vertex corrections. These
effects are model dependent and can be sensitive to the values of some
parameters such as fermion and Higgs masses. So far the SM seems to be
consistent with the precision electroweak measurements
and obviously any other viable model should also agree
with the data. We note, as discussed in \cite{4GenLet}, the data
impose two important constraints on new fermion $SU(2)$ doublets
in our models:
\begin{enumerate}
\item The mass squared differences within any new fermion
$SU(2)$ doublets must be small ($\ll (100 \; \rm{GeV})^2$),
in order that the predicted value of the $\rho$
parameter should not deviate too much from its experimental
value close to unity.
\item The number of new $SU(2)$ doublets is severely restricted
by the measured value of the S parameter or its equivalent
\cite{PesAlNov}.
\end{enumerate}

%and there is no
%experimental evidence that the SM should not be correct.
%Obviously any other viable
%model should also agree with the data and we discuss the radiative corrections
%for our model in \cite{4GenLet}. We simply note here that we take the mass
%squared differences within any new fermion $SU(2)$ doublets to be small
%($\ll (100 \; \rm{GeV})^2$), in order that the predicted value of the $\rho$
%parameter should not deviate too much from its experimental value close to
%unity.

%The parameters used to describe the loop corrections are called
%$\epsilon_{1,2,3}$ \cite{Epsilons} or $S, T, U$ \cite{STU}. More detailed
%parameterisations are possible but these cannot be accurately calculated for
%technicolour-type models (among others) and so are of limited use in
%comparing the SM to most alternatives.
%
%A detailed analysis of these parameters is given in \cite{4GenLet} but in this
%paper we do not examine them.
%The main constraint comes from the fact that the parameter
%\begin{displaymath}
%\rho \equiv \frac{M_W^2}{M_Z^2 cos^2 \theta_W }
%\end{displaymath}
%is experimentally determined to be close to one. The deviation of $\rho$ from
%one is the parameter $\epsilon_1$ and so we must have $\epsilon_1 \approx 0$.
%This means that the differences of the mass squared of two fermions in the
%same $SU(2)$ doublet should be small ($\ll (100 \rm{GeV})^2$).

%*************

\section{Fermion Mass and Anomaly Cancellation}
\label{Simplifications}

In the SM fermions get a mass via the Higgs mechanism. To do this in a general
gauge group of the form
\begin{displaymath}
U(1) \otimes SU(2) \otimes G/D
\end{displaymath}
where $G$ is any Lie group and $D$ is a discrete group, using the SM Higgs
particle, a left-handed fermion representation $(y,{\bf 2},{\bf R})$
should occur together with the left-handed anti-fermion representations
$(-[y+1],{\bf 1},{\bf \overline{R}})$ and $(-[y-1],{\bf 1},
{\bf \overline{R}})$. We shall refer to this
as the mass grouping $\{y,{\bf R}\}$ where {\bf R} is an
irreducible representation of $G$.
As explained in section~\ref{masslessfermions} we assume that all fermions in
our models, other than the leptons which have already been observed, get a mass
by this mechanism.
We shall now describe what consequences this has for anomaly cancellation in
our models, where $G$ is a product of $SU(N_i)$ groups with $N_i \ge 3$.

%It therefore seems reasonable to assume that all additional
%fermions will be grouped as above in order to become massive, though it is
%not absolutely necessary
%provided that any extra massless fermions are not in a singlet representation
%of some $SU(N)$ group for any $N>3$. This is because the only reason any
%massless particles would not have already been seen is if
%the confinement scale
%of $SU(N)$ was higher than current accelerator energies.

We consider the grouping $\{y,{\bf R}\}$ for the gauge group
\begin{displaymath}
U(1) \otimes SU(2) \otimes \Pi_i SU(N_i)
\end{displaymath}
where the irreducible representation {\bf R} is made up of
fundamental (${\bf N}_i$ or
${\bf \overline{N}}_i$) or singlet representations of each
factor $SU(N_i)$. The
contribution to each type of anomaly from this grouping,
$\{y,{\bf R}\}$, is easily
calculated, using the results of section~\ref{Anoms}, to be as follows.
\begin{displaymath}
\begin{array}{lllll}
[SU(N_{i})]^{3} & \rightarrow & 2S_{R}n+S_{R}(-n)+S_{R}(-n) & = & 0 \\
\left[SU(N_{i})\right]^{2}U(1) & \rightarrow & 2S_{R}n^{2}y-S_{R}n^{2}(y+1)
-S_{R}n^{2}(y-1) & = & 0 \\
\left[{\rm Grav}\right]^{2}U(1) & \rightarrow & 2S_{R}y+S_{R}(-y-1)
+S_{R}(-y+1) & = & 0 \\
\left[U(1)\right]^{3} & \rightarrow & 2S_{R}y^{3}+S_{R}(-y-1)^{3}
+S_{R}(-y+1)^{3} & = & -6S_{R}y \\
\left[SU(2)\right]^{2}U(1) & \rightarrow & & & 2S_{R}y
\end{array}
\end{displaymath}
Here $n_i$ is the $N_i$-ality of the representation {\bf R} and $S_R$ is its
dimension (size).

So we can see that the above grouping which is necessary to give a mass to the
fermions also simplifies the anomaly constraints. In particular, if we take
all fermions to be grouped in this way then we are only left with the single
constraint for the absence of the mixed gauge-gravitational and gauge anomalies
\begin{equation}
\label{AllAnoms}
\sum_{j}S_{j}y_{j}=0
\end{equation}
where j labels each grouping $\{y_j, {\bf R}_j\}$.

There will also be no Witten anomaly, since we must have an even number of
$SU(2)$ doublets to satisfy eq.~(\ref{AllAnoms}). This follows from the
charge quantisation rule eq.~(\ref{GenChQuant}), the fact that $N_{i}$ are all
odd and the assumption of fundamental or singlet representations for each
$SU(N_i)$ subgroup. Using the charge quantisation rule and defining
\begin{equation}
\frac{e_j}{d_j} = \sum_i \frac{m_{N_i}}{N_i} (n_i)_j
\end{equation}
we can write
\begin{equation}
\frac{y_{j}}{2} = c_{j} + \frac{1}{2} +\frac{e_{j}}{d_{j}}
\end{equation}
where $c_{j}, d_{j}$ and $e_{j}$ are integers and $d_{j}$ are odd. Therefore,
since eq.~(\ref{AllAnoms}) can be written as
$\sum_{j}S_{j}\frac{y_{j}}{2}=0$,
we must have $\sum_{j}S_{j}\frac{1}{2} \equiv 0 \pmod 1$. In other words
$\sum_{j}S_{j} \equiv 0 \pmod 2$, which means that there are an even number of
$SU(2)$ doublets and so no Witten anomaly.

\section{The $SMG_{235}$ Model Without New SM Fermions}
\label{Only5sFails}

%We could try and argue that all fermions which were SU(5) singlets would be SM
%particles and would already have been observed since it is unlikely that they
%would have a mass significantly higher than the top mass. However, as we shall
%argue in section~\ref{SU2*SUN}, a fourth generation of quarks
%(but not leptons)
%with a mass slightly higher than the top mass cannot yet be ruled out by
%experiment.

%***************
%\input{only5s}
%
%***********************
%* Start of Only5s.tex *
%***********************
%

Here we will examine the model based on the gauge group
$SMG_{235} \equiv G_5$ defined in
eqs.~(\ref{SMG235Group}) and (\ref{SMG235Discrete}), since it is the absolute
minimal extension to the SM among all the possible groups we have proposed in
section~\ref{Groups}. In section~\ref{SU2*SUN} we will consider models based on
the groups $SMG_{23N}$ of eqs.~(\ref{SMG2NNGroup}) and (\ref{SMG2NNDiscrete}),
including new SM fermions to highlight the general features of all such
extensions to the SM. However we will only analyse the consequences in detail
for $SMG_{235}$.

In this section we will discuss the two possibilities: (i) that there are no
new fermions beyond those of the SM and (ii) that there are new fermions which
all couple to the $SU(5)$ gauge group. This latter possibility may seem to be
tantamount to adding a completely separate sector to the SM rather than
extending the SM, since the new fermions will be confined under a new gauge
group. However, it is really no more a separate sector than the SM
is three separate sectors (one for each generation), since these extra fermions
will still couple to the electro-weak group due to the charge quantisation
rule. We will discuss the other
possibility, that there are new fermions, some
coupling to the $SU(5)$ gauge group and others not, in section~\ref{SU2*SUN}

\subsection{No New Fermions}

There is of course the possibility that there are no extra fermions associated
with this enlarged group. If this is so then the only possible
observations would be the detection of $SU(5)$ `glueballs'. In this case the
$SU(5)$ gauge group would be decoupled from the $SMG$ and so the only way to
observe the glueballs would be through their gravitational interactions. They
could have been produced in the very early universe and the lightest state
would be essentially stable since they could only decay via the gravitational
interaction. Therefore they would only be observable as dark matter.

%This is a possible source of dark matter in the universe. However we would
%%have no way
%of predicting the mass of the lightest glueballs in this case since there
%%would
%be no way of measuring the strength of the $SU(5)$ gauge coupling. All that
%could be done would be to fit the mass to experimental data but this is
%%unlikely
%to distinguish this possibility from other weakly interacting massive
%%particles
%as candidates for cold dark matter.
%
%Therefore we shall examine the case that there are more fermions than have
%currently been
%observed and we shall show whether or not we can construct a consistent model.

So this case is essentially uninteresting and will not be considered further.
Instead we turn to the possibility that there exist more types of fermions than
have been currently observed and consider whether or not they can be
incorporated into a consistent model.

\subsection{New Fermions Coupling to $SU(5)$}
\label{5sOnlySoln}

Of course fermions all contribute to anomalies which must be
cancelled. The fermions in the SM cancel all anomalies on their
own; so the extra fermions must cancel all anomalies amongst
themselves.

%We shall examine all possible combinations of fundamental representations
%which are not $SU(5)$ singlets without making the assumption that all fermions
%must get a mass via the Higgs field as discussed in
%section~\ref{Simplifications}. We will proceed with this approach as far as
%possible before examining the solution when all additional fermions are
%massive.

As explained in section~\ref{Simplifications} the anomaly equations in our
models are greatly
simplified when all the fermions are massive due to the SM Higgs mechanism. In
fact they are reduced to just one equation, $\sum_{i}S_{i}y_{i}=0$.
If we label each mass-grouping of fermion representations by the label
$\{y,{\bf R}\}$ where {\bf R} is the representation of the group $SU(3) \otimes
SU(5)$,
then table~\ref{SU5Reps} shows all six possible groupings, $a$ to $f$, and
their
relative contributions, $S_{i}y_{i}$, to the anomaly equation. We use
eq.~(\ref{ChQuant235})
with the definition $m \equiv m_5$ to simplify the notation, giving us the
charge quantisation rule,
\begin{equation}
\label{ChQuantSU5}
\frac{y}{2}+\frac{1}{2}\rm{``duality''}+\frac{1}{3}\rm{``triality''}
+\frac{m}{5}\rm{``quintality''} \equiv 0 \pmod 1
\end{equation}
where the integer m is fixed in any given model
\footnote{In fact we can limit m to be
1 or 2 since it is only defined modulo 5 and, by replacing $m$ with
$-m \pmod 5$ and all representations of $SU(5)$ with their conjugates, we are
left with an equivalent model.}.
So we can determine $\frac{y}{2} \pmod 1$ for any given representation {\bf R}.

\begin{table}[htb]
\caption{Allowed mass groupings $\{y,{\bf R}\}$ of new fermions in the
$SMG_{235}$
model, using the charge quantisation rule,
eq.~(\protect\ref{ChQuantSU5}\protect), and fundamental representations
of $SU(5)$. Their relative contributions to the anomaly equation,
eq.~(\protect\ref{AllAnoms}\protect), are given in the final column. A
particular mass grouping of type $t$ is given by choosing a particular value of
weak hypercharge, i.e.\ by choosing a particular value of the integer $N_t$.}
\begin{displaymath}
\begin{array}{|c|c|c|c|} \hline
$Type$ & {\bf R} & \frac{y}{2} & \frac{1}{10}Sy \\ \hline
a & {\bf 1},{\bf 5} & N_a-\frac{m}{5}-\frac{1}{2} & N_a-\frac{m}{5}-\frac{1}{2}
\\ \hline
b & {\bf 1},{\bf \overline{5}} & N_b+\frac{m}{5}+\frac{1}{2} &
			N_b+\frac{m}{5}+\frac{1}{2} \\ \hline
c & {\bf 3},{\bf 5} & N_c-\frac{m}{5}+\frac{1}{6} &
3N_c-\frac{3m}{5}+\frac{1}{2} \\ \hline
d & {\bf 3},{\bf \overline{5}} & N_d+\frac{m}{5}+\frac{1}{6} &
			3N_d+\frac{3m}{5}+\frac{1}{2} \\ \hline
e & {\bf \overline{3}},{\bf 5} & N_e-\frac{m}{5}-\frac{1}{6} &
			3N_e-\frac{3m}{5}-\frac{1}{2} \\ \hline
f & {\bf \overline{3}},{\bf \overline{5}} & N_f+\frac{m}{5}-\frac{1}{6}
			& 3N_f+\frac{3m}{5}-\frac{1}{2} \\ \hline
\end{array}
\end{displaymath}
\label{SU5Reps}
\end{table}

For a solution to the anomaly equation $\sum_i S_iy_i = 0$, we must obviously
combine the fractions $\frac{m}{5}$ so that the
$5$ is cancelled in the denominator since all $N$s are integers. We must also
have an even number of groupings so that the $\frac{1}{2}$ 's combine to give
an integer. This automatically ensures that there can be no Witten anomaly as
explained in section~\ref{Simplifications}.
This can be done by using equal numbers of type $a$ and type $b$ groupings.
The two smallest
solutions are in fact: (i) one type $a$ grouping and one type $b$ grouping and
(ii) two groupings of type $a$ and two of type $b$.
The smallest solution, (i), is not possible without giving the fermions
a fundamental Dirac mass, since the anomaly constraints require that
$N_{a}+N_{b}=0$ giving pairs of representations, $(y,{\bf 2},{\bf 1},{\bf 5})$
and
$(-y,{\bf 2},{\bf 1},{\bf \overline{5}})$ etc., which are not mass protected.

The smallest allowed solution with mass protected fermions is therefore
solution (ii) with two groupings of type $a$ and two of type $b$.
This solution is shown in detail in Table~\ref{Only5sSoln}.
All anomalies cancel provided $\sum_{i=1}^4 N_i=0$.
We can now choose values of the $N_i$.
\begin{table}
\caption{Smallest anomaly-free (subject to the constraint $N_1+N_2+N_3+N_4=0$)
set of mass protected fermions which all couple to $SU(5)$.}
\begin{displaymath}
\begin{array}{|c|c|} \hline
\rm{Representation \; under} & U(1) \rm{\; Representation} \\
SU(2) \otimes SU(3) \otimes SU(5) & \frac{y}{2} \\ \hline
{\bf 2},{\bf 1},{\bf 5} & N_1-\frac{m}{5}-\frac{1}{2} \\ \hline
{\bf 1},{\bf 1},{\bf \overline{5}} & -N_1+\frac{m}{5} \\ \hline
{\bf 1},{\bf 1},{\bf \overline{5}} & -N_1+\frac{m}{5}+1 \\ \hline
{\bf 2},{\bf 1},{\bf 5} & N_2-\frac{m}{5}-\frac{1}{2} \\ \hline
{\bf 1},{\bf 1},{\bf \overline{5}} & -N_2+\frac{m}{5} \\ \hline
{\bf 1},{\bf 1},{\bf \overline{5}} & -N_2+\frac{m}{5}+1 \\ \hline
{\bf 2},{\bf 1},{\bf \overline{5}} & N_3+\frac{m}{5}+\frac{1}{2} \\ \hline
{\bf 1},{\bf 1},{\bf 5} & -N_3-\frac{m}{5}-1 \\ \hline
{\bf 1},{\bf 1},{\bf 5} & -N_3-\frac{m}{5} \\ \hline
{\bf 2},{\bf 1},{\bf \overline{5}} & N_4+\frac{m}{5}+\frac{1}{2} \\ \hline
{\bf 1},{\bf 1},{\bf 5} & -N_4-\frac{m}{5}-1 \\ \hline
{\bf 1},{\bf 1},{\bf 5} & -N_4-\frac{m}{5} \\ \hline
\end{array}
\end{displaymath}
\label{Only5sSoln}
\end{table}

The fermion contribution to the (first order)beta function for the $U(1)$
running gauge coupling constant is proportional to $\sum y^2$. We therefore
want to choose values of $N_i$ so as to minimise $\sum y^2$, in order that
any $U(1)$ Landau pole is at as high an energy as possible.
This gives us the best chance that the solution of table~\ref{Only5sSoln}
will be perturbatively valid up to the Planck scale and hence that our model
will be self-consistent. However, this condition of minimising $\sum y^2$ is
also suggested by the small representation structure of the SM,
as explained in section~\ref{SmallReps}. Keeping in mind that the $N_i$
are integers, $\sum_{i=1}^4 N_i=0$, and that the particles must be mass
protected, we find that the minimum value of $\sum y^2$ is given by
\begin{displaymath}
\begin{array}{ccc}
N_1=N_2=1 & N_3=0 & N_4=-2 \\
 & $or$ & \\
N_3=N_4=-1 & N_1=0 & N_2=2
\end{array}
\end{displaymath}
where $m=2$. These values of $N_i$ give $\sum y^2 = 203.2$, for the
solution of table~\ref{Only5sSoln}, which is much
larger than the $\frac{40}{3}$ per generation of the SM particles.

%We must now try and estimate the mass scale of the additional particles in
%order to see how they effect the running coupling constants. In particular
%we would like an upper limit on the masses since this would show the minimum
%effect such particles could have on the running coupling constants. Once we
%have estimated an upper bound on the particle masses we can see whether it
%is possible to have these particles without having any Landau poles below the
%Planck scale.

%In section~\ref{Intro} we explained that in this model the $SU(5)$ gauge
%group would act as a technicolour and generate some of the mass of the
%$W^{\pm}$ and $Z^{0}$ bosons.
In section~\ref{TechniColour} we explained that it was reasonable to consider
that all new fermions could be included at a threshold no higher than 1.7 TeV.
This should provide an accurate enough upper limit for the threshold for our
purposes. Therefore, since the fermions will have the least effect on the
running coupling constants if they are included at the highest possible
threshold, we will assume that all these extra fermions can be included with a
simple threshold at 1.7 TeV. We can now check whether or not this model has a
Landau pole below the Planck scale.

There are four fine structure constants which we shall label by
$\alpha_{1}$, $\alpha_{2}$, $\alpha_{3}$ and $\alpha_{5}$ corresponding to
the four gauge groups $U(1)$, $SU(2)$, $SU(3)$ and $SU(5)$ respectively. The
fine structure constants, $\alpha_{i}$, are related to the gauge coupling
constants,
$g_{i}$, by the relation $\alpha_{i}=\frac{g_{i}^{2}}{4\pi}$. The equations
governing the running coupling constants to first order in perturbation theory
\cite{RGEs1} (a good discussion of RGEs in the SM is given in \cite{SMRGE}) can
be integrated analytically to give
\begin{eqnarray}
\label{runalpha1SM}
\frac{1}{\alpha_{1}(\mu)} & = & \frac{1}{\alpha_{1}(\mu_{0})}-\frac{1}{12\pi}
\left(Y^{2}+n_{H}\right)ln\left(\frac{\mu}{\mu_{0}}\right) \\
\label{alpha2}
\frac{1}{\alpha_{2}(\mu)} & = & \frac{1}{\alpha_{2}(\mu_{0})}+\frac{1}{12\pi}
\left(44-2n_{2f}-n_{H}\right)ln\left(\frac{\mu}{\mu_{0}}\right) \\
\label{alpha3}
\frac{1}{\alpha_{3}(\mu)} & = & \frac{1}{\alpha_{3}(\mu_{0})}+\frac{1}{12\pi}
\left(66-2n_{3f}\right)ln\left(\frac{\mu}{\mu_{0}}\right) \\
\label{alpha5}
\frac{1}{\alpha_{5}(\mu)} & = & \frac{1}{\alpha_{5}(\mu_{0})}+\frac{1}{12\pi}
\left(110-2n_{5f}\right)ln\left(\frac{\mu}{\mu_{0}}\right)
\end{eqnarray}
where we calculate $\alpha_i(\mu)$ (the running coupling constants at the
energy scale $\mu > \mu_0$) in terms of $\alpha_{i}(\mu_{0})$.
$Y^{2} \equiv \sum y^2$ is the sum of the weak hypercharges squared for all
fermions with a mass below $\mu_{0}$ and $n_{mf}$ are the number of fermion
{\bf m}
and ${\bf \overline{m}}$ representations of $SU(m)$ with mass below $\mu_{0}$.
$n_{H}$
is the number of Higgs doublets with mass below $\mu_{0}$. These equations
assume that there are no fermions or Higgs scalars with a mass between
$\mu_{0}$ and $\mu$. In order to calculate the value of $\alpha_{i}(\mu)$
when there are fermions or Higgs bosons with masses between $\mu_{0}$ and
$\mu$ we must do the calculation in steps, calculating the value of
$\alpha_{i}$ up to the mass of each particle. So we use the experimental values
of the fine structure constants at $M_Z$ (including the top quark and Higgs
boson in the beta functions at this scale) to calculate
the coupling constants at 1.7 TeV, where we include the new fermions, and then
run the coupling constants up to the Planck scale. This is a crude method since
there would really be complicated threshold effects as each fermion was
included. However these effects can reasonably be assumed to be small, relative
to the changes in the coupling constants caused by the running from the
electroweak scale to the Planck scale, and so we will use this much simpler
method. Second order RGEs \cite{RGEs2} could be used but the improvement over
the first order RGEs would not be significant when compared to the error
introduced by the naive assumptions made about threshold effects.

{}From \cite{PDG} we find
\begin{eqnarray}
\label{alpha1SM}
\alpha_{1}^{-1}(M_{Z}) & = & 98.08 \pm 0.16 \\
\alpha_{2}^{-1}(M_{Z}) & = & 29.794 \pm 0.048 \\
\alpha_{3}^{-1}(M_{Z}) & = & 8.55 \pm 0.37
\end{eqnarray}

\begin{figure}
\epsfxsize=\textwidth
\epsffile[100 100 500 500]{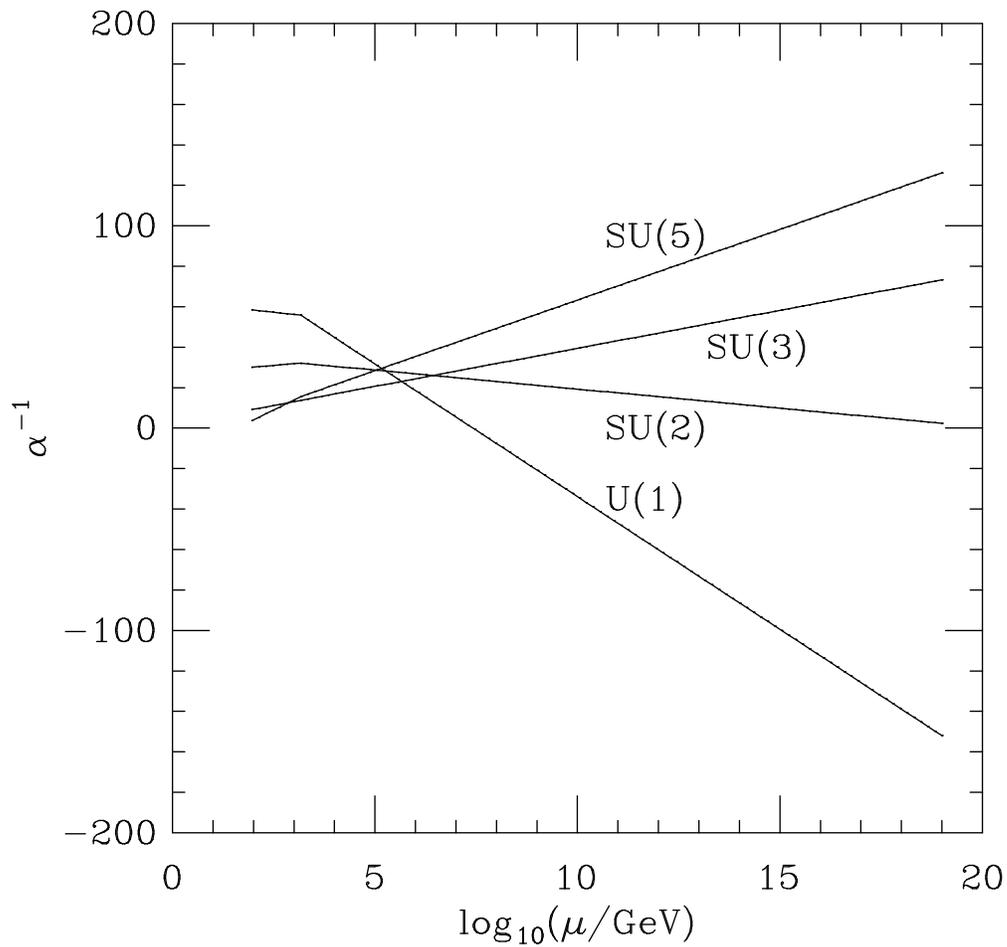}
\caption{$\alpha^{-1}$ from $M_Z$ to the Planck scale for each component group
in the $SMG_{235}$ model without new SM fermions. There is
clearly a $U(1)$ Landau pole at $\mu \sim 10^7$ GeV and $SU(2)$ also loses
asymptotic freedom. $\alpha^{-1}_{5}(M_Z)=2$ has been chosen as a specific
example.}
\label {alph5s}
\end{figure}

We can now use the above equations to examine how the coupling constants behave
up to the Planck scale. Since there is no experimental value for $\alpha_5$ at
any energy scale we shall assume that $\alpha_5^{-1}(M_Z) = 2$, so that the
$SU(5)$ interaction is stronger than QCD at $M_Z$ and confines at the
electroweak scale. Fig.~\ref{alph5s} shows what happens for each group. For the
graphs, we normalise the $U(1)$ gauge coupling as if the $U(1)$ group was
embedded in a simple group. This essentially corresponds to a redefinition of
$g_1$:
\begin{eqnarray}
(g_1^2)_{\rm{GUT}} & \equiv & \frac{5}{3}(g_1^2)_{\rm{SM}} \\
(\alpha_1^{-1})_{\rm{GUT}} & \equiv & \frac{3}{5}(\alpha_1^{-1})_{\rm{SM}}
\end{eqnarray}

So henceforth we use the standard GUT normalisation.
Eqs.~(\ref{runalpha1SM}) and (\ref{alpha1SM}) now become,
\begin{eqnarray}
\label{runalpha1GUT}
\frac{1}{\alpha_{1}(\mu)} & = & \frac{1}{\alpha_{1}(\mu_{0})}-\frac{1}{20\pi}
\left(Y^{2}+n_{H}\right)ln\left(\frac{\mu}{\mu_{0}}\right) \\
\label{alpha1GUT}
\alpha_1^{-1}(M_Z) & = & 58.85 \pm 0.10
\end{eqnarray}
%
%**************
%**************

As we can see from Fig.~\ref{alph5s}, $\frac{1}{\alpha_{1}}$ becomes negative
at about $10^7$ GeV
which means that there is a Landau pole. So we can conclude that this theory
would be inconsistent, at least as far as perturbation theory is concerned,
without new interactions below $10^7$ GeV.

In fact we can show that there is no anomaly-free model, having all new
fermions coupling to $SU(5)$, with a desert above
the TeV scale, which does not have a Landau pole below the Planck scale.
The condition for no Landau pole below the Planck scale is
$\frac{1}{\alpha_{1}(M_{\rm Planck})}>0$. Therefore eq.~(\ref{runalpha1GUT})
can be rearranged to give
\begin{equation}
\label{Y^2max}
Y^{2}+n_{H} \qquad < \qquad
\frac{20\pi}{\alpha_{1}(\mu_{0})ln\left(\frac{M_{\rm Planck}}{\mu_{0}}
						\right)}
\end{equation}

Since, for the SM, $Y^{2}_{SM}=40$ and $n_H=1$,
\begin{equation}
Y^2+n_H \qquad \ge \qquad 41
\end{equation}
above the electroweak scale and so we can use
eqs.~(\ref{runalpha1GUT})~and~(\ref{alpha1GUT}) to
calculate an upper limit for $\frac{1}{\alpha_{1}(1.7 \; {\rm TeV})}$:
\begin{equation}
\frac{1}{\alpha_{1}(1.7\; {\rm TeV})} \qquad \le \qquad 57
\end{equation}
We then use $Y^2=Y^2_{SM}+
Y^2_{\rm new}$ in eq.~(\ref{Y^2max}) with $\mu_0=1.7$ TeV and conclude that
\begin{equation}
\label{MaxY2}
Y^{2}_{\rm new} \qquad < \qquad 57.5
\end{equation}
assuming the new fermions can be included naively at a threshold no higher
than 1.7 TeV.

For each mass grouping $\{y,{\bf R}\}$, with $P_R = 4S_R$ fermions, we can
calculate
the value of $Y^{2}$:
\begin{equation}
Y^{2}=S_{R}[2y^{2}+(y+1)^{2}+(y-1)^{2}]=S_{R}(4y^{2}+2)
\end{equation}
Therefore we have,
\begin{equation}
\label{MinY2}
Y^2 \quad \ge \quad 2S_R \quad = \quad \frac{1}{2}P_R
\end{equation}
If there are several mass groupings,
$Y^2 \ge \frac{1}{2} \sum P_R \equiv \frac{1}{2} P$ where P is the total number
of fermions. So if we define $P_{\rm new}$ to be the number of non-SM fermions,
we can conclude,
\begin{equation}
P_{\rm new} \quad \le \quad 2Y^{2}_{\rm new} \quad < \quad 115
\end{equation}

So now we have shown that there must be less than 115 extra fermions.
However the smallest solutions, subject to the constraints in this section,
larger than two type $a$ and two type $b$ representations are three type $a$
and one type $c$ representations etc.\ which
contain 120 fermions and so must cause a Landau pole below the Planck scale
\footnote{Using second order RGEs or a more complete analysis of thresholds
would obviously change the precise limit in eq.~(\ref{MaxY2}). However, the
charge quantisation rule in our model means that $y$ cannot be zero and so
it is not possible to attain the limit of eq.~(\ref{MinY2}). In fact, the
value of $Y_{\rm new}^2$ will generally be much greater than this limit. For
example, three type $a$ and one type $c$ lead to $Y_{\rm new}^2 \ge
\frac{1708}{15}
\approx 114$ which is much greater than the required maximum given by
eq.~(\ref{MaxY2}).}.
Therefore there are no possible anomaly-free models without a Landau pole,
where
all the new fermions couple to the $SU(5)$ gauge group.

We will now examine the case where we allow some new $SU(5)$ singlet fermions,
as well
as some fermions which couple to $SU(5)$, in order to cancel the anomalies.
We shall show that it is possible to have more SM fermions in such a model.

%
%
%*********************
%* End of Only5s.tex *
%*********************
%

%***************

\section{The $SMG_{235}$ Model With New SM Fermions}
\label{SU2*SUN}

In this section we shall first examine sets of fermions (which are
generalisations of the SM quarks) in groups, defined by
eqs.~(\ref{SMG2NNGroup}) and (\ref{SMG2NNDiscrete}), similar to the
$SMG$. We shall then examine the
particular case of the group $SMG_{235}$ and discuss the possibility of
experimental evidence for and against this self-consistent model.

\subsection{Fermions in the groups $SMG_{2M}$ and $SMG_{2MN}$}
\label{SMG2M+2MN}

In section~\ref{SMG2M} we shall examine the group $SMG_{2M}$. The $SMG$ is an
example of this type of group where $M=3$. We shall show that
this general group
allows anomaly-free sets of fermions which consist of a
generation of SM leptons
and a generation of $SU(M)$ `quarks' which are a simple generalisation of the
$SU(3)$ quarks in the SM\@.

We shall then show in section~\ref{SMG2MN} that we can have anomaly-free sets
of fermions in the group $SMG_{2MN}$ without any leptons. We shall then examine
the particular case of the group $SMG_{235}$ which we shall discuss in detail.

\subsubsection{Fermions in the Group $SMG_{2M}$}
\label{SMG2M}

In the SM, each generation is formed by taking the two mass groupings
$\{\frac{1}{3},{\bf 3}\}$ and $\{-1,{\bf 1}\}$ (where the
representations {\bf 3} and {\bf 1} are of the group
$SU(3)$) as explained in section~\ref{Simplifications}
and appendix~\ref{SMGen}.
We will now consider a more general situation where we have the gauge group
$SMG_{2M}$ defined in section~\ref{Groups} (where $M>2$ is a prime number) and
the fermions are in the groupings $\{y_{1},{\bf M}\}$ and
$\{y_2,{\bf 1}\}$ (where the
representations {\bf M} and {\bf 1} are of the group $SU(M)$).

{}From section~\ref{Simplifications} all the gauge anomalies will cancel if
\begin{equation}
My_{1}+y_{2}=0
\end{equation}
Since we also have the charge quantisation rule
\begin{equation}
\frac{y}{2}+\frac{1}{2}\rm{``duality"}+\frac{m_{M}}{M}
\rm{``M-ality"} \equiv 0 \pmod{1}
\end{equation}
we can write
\begin{eqnarray}
\frac{y_{1}}{2} & = & -\frac{1}{2}-\frac{m_{M}}{M}+c_{1} \\
\frac{y_{2}}{2} & = & -\frac{1}{2}+c_{2}
\end{eqnarray}
where $c_{1}$ and $c_{2}$ are integers.
We now have the condition that for no anomalies to be present
\begin{equation}
-\frac{M+1}{2}-m_M+Mc_{1}+c_{2}=0
\end{equation}

In the SM a lepton generation is formed (with the addition of a right-handed
neutrino which can be removed without effecting any anomalies) when we have
$c_{2}=0$ as explained
in appendix~\ref{SMGen}. If we
insert this value into the above equation then we find
\begin{equation}
c_{1}=\frac{1}{M}\left(\frac{M+1}{2}+m_{M}\right)
\end{equation}
This can always be solved by setting $m_{M}=\frac{M-1}{2}$. In fact if $M=3$
then this is simply one of the anomaly-free SM quark-lepton generations.

However, this is not a good solution for an extension of the SM (which would be
obtained by considering $SMG_{2M} \subset SMG_{23M}$) since it contains an
extra massless neutrino which has already been ruled out by experiment.
It is difficult to produce a neutrino with a mass so
large that it wouldn't already have been detected,
%since we cannot simply add
%a right-handed neutrino
as explained in section~\ref{expmass}. We
could choose not to set $c_{2}=0$ or 1 above, which would force all the extra
leptons to be massive (by leptons we mean any fermions which are only coupled
to the electroweak subgroup, $SU(2) \otimes U(1)$). This is because there would
then be two $SU(2)$ singlets which were charged (and at least one would have an
electric charge of two or more, which is against our principle of small
representations) and so both could get a mass
by the usual SM Higgs mechanism since neither could get a Majorana mass.
%However, we would expect that these leptons would already have been seen since
%there is no natural explanation why they would get a mass much larger than
%can be detected by present experiments yet the known leptons have a mass much
%smaller than the limits of present experiments.
But even if we assumed that
these leptons had masses higher than experimental limits this solution is not
really favoured by our postulate of small values of weak hypercharge discussed
in section~\ref{SmallReps}. So in order to find a satisfactory
solution we shall look at a similar general case.

\subsubsection{Fermions in the Group $SMG_{2MN}$}
\label{SMG2MN}

Suppose we have the gauge group $SMG_{2MN}$,
where both $M$ and $N>M\ge 3$ are mutually prime integers, which has the charge
quantisation rule
\begin{equation}
\frac{y}{2}+\frac{1}{2}\rm{``duality"}+\frac{m_{M}}{M}\rm{``M-ality"}
+\frac{m_{N}}{N}\rm{``N-ality"}
\equiv 0 \pmod{1}
\end{equation}
Then with fermions in mass groupings $\{y_{1},({\bf M},{\bf 1})\}$ and
$\{y_{2},({\bf 1},{\bf N})\}$ (where the representations $({\bf M},{\bf 1})$
and $({\bf 1},{\bf N})$ are of the group $SU(M) \otimes SU(N)$) the condition
for no anomalies is
\begin{equation}
My_{1}+Ny_{2}=0
\end{equation}
The charge quantisation rule means that we can write
\begin{eqnarray}
\frac{y_{1}}{2} & = & -\frac{1}{2}-\frac{m_{M}}{M}+c_{1} \\
\frac{y_{2}}{2} & = & -\frac{1}{2}-\frac{m_{N}}{N}+c_{2}
\end{eqnarray}
where $c_{1}$ and $c_{2}$ are integers. We then find that the
condition for no anomalies becomes
\begin{equation}
2Nc_{2}=N+[2(m_M+m_N)+(1-2c_{1})M]
\end{equation}
Both $N$ and $M$ are odd and therefore there will always be
a solution, since we can
choose $(m_{M}+m_{N})=M$ and $3-2c_{1}$ to be an odd multiple of $N$. In
general there will also be other solutions.
%
%\begin{equation}
%[2(m_M+m_N)+(1-2c_{1})M] \equiv 0 \pmod{N}
%\end{equation}
%This is always possible since we can choose $(m_{M}+m_{N})=M$ and
%$3-2c_{1}$ to be an odd multiple of $N$. (In general there will also be other
%solutions).

In particular, for the gauge group $G_{5} \equiv SMG_{235}$
we can have a fourth generation of quarks without any extra
leptons, by choosing $M=3$, $N=5$, $m_{3}=1$ and $c_{1}=1$ above. Then
\begin{equation}
10c_{2}=5+[2(1+m_{5})-3]
\end{equation}
or equivalently
\begin{equation}
5c_{2}=2+m_{5}
\end{equation}
So we have a solution with $c_{2}=1$ and $m_{5}=3$.

The representations of the left-handed fermions which couple to the $SU(5)$
subgroup are shown in Table~\ref{SU5Quarks}. This is a generalisation of
the quarks in the SM, coupling to $SU(5)$ rather than $SU(3)$.

%If we set
%$N=3$ we would in fact get a generation of quarks with the opposite chirality
%to those in the SM. This is not surprising since we are using these fermions
%to cancel the anomaly contribution of a 4th generation of SM quarks.
%
\begin{table}
\caption{Left-handed fermions coupling to $SU(5)$ in the mass grouping
$\{-\frac{1}{5},({\bf 1},{\bf 5})\}$. The electric charges are
in units of $\frac{1}{5}$ due to the charge quantisation rule.}
\begin{displaymath}
\begin{array}{|c|c|c|}
\hline
\rm{Representation\;under} & U(1)\;\rm{Representation} &
					\rm{Electric\;Charge} \\
SU(2) \otimes SU(3) \otimes SU(5) & \frac{y}{2} & Q \\ \hline
2,1,5 & -\frac{1}{10} & \left ( \begin{array}{c} \frac{2}{5} \\
				 -\frac{3}{5} \end{array} \right ) \\ \hline
1,1,\overline{5} & -\frac{4}{10} & -\frac{2}{5} \\ \hline
1,1,\overline{5} & \frac{6}{10} & \frac{3}{5} \\ \hline
\end{array}
\end{displaymath}
\label{SU5Quarks}
\end{table}

In fact we have a solution with a fourth generation of quarks for the general
group $SMG_{23N}$, where $N$ is any odd integer
greater than but not divisible by 3, by choosing $c_{2}=1$
and $m_{N}=\frac{1}{2}(N+1)$. This means that, if a fourth generation of quarks
without leptons was detected, there would be no immediate way of deducing the
value of $N$. Table~\ref{SUNQuarks} shows the properties of the left-handed
fermions which couple to the $SU(N)$ subgroup. Note that this is a
generalisation of the SM quarks, coupling to $SU(N)$ with the specific choice
of $m_N=\frac{1}{2}(N+1)$. If we set $N=3$ we would in fact get a generation
of quarks with the opposite chirality to those in the SM. This is to be
expected since we are using these fermions to cancel the anomaly contribution
of a 4th generation of SM quarks (with the usual chirality).
\begin{table}
\caption{Fermions coupling to $SU(N)$ which would form an anomaly-free set of
fermions together with a fourth generation of quarks.}
\begin{displaymath}
\begin{array}{|c|c|c|}
\hline
\rm{Representation\;under} & U(1) \;\rm{Representation} &
					\rm{Electric\;Charge} \\
SU(2) \otimes SU(3) \otimes SU(N) & \frac{y}{2} & Q \\ \hline
{\bf 2},{\bf 1},{\bf N} & -\frac{1}{2N} & \left ( \begin{array}{c}
\frac{N-1}{2N} \\
				-\frac{N+1}{2N} \end{array} \right ) \\ \hline
{\bf 1},{\bf 1},{\bf \overline{N}} & -\frac{N-1}{2N}
& -\frac{N-1}{2N} \\ \hline
{\bf 1},{\bf 1},{\bf \overline{N}} & \frac{N+1}{2N} & \frac{N+1}{2N} \\ \hline
\end{array}
\end{displaymath}
\label{SUNQuarks}
\end{table}

This solution, with a fourth generation of quarks and the fermions of
Table~\ref{SU5Quarks}, for the gauge group $SMG_{235}$ is analogous to one
SM quark-lepton generation in the gauge group
$SMG$, in the sense that it is the smallest anomaly-free set of
mass-protected fermions which
couple non-trivially to all the gauge fields. The SM quark-lepton generation is
shown to be the smallest such set of fermions for the $SMG$ in
appendix~\ref{SMGen}. Note
that although a generation of SM leptons and the fermions conjugate to those in
Table~\ref{SU5Quarks} is a smaller anomaly-free set of fermions in the gauge
group $SMG_{235}$, none of these fermions couples to the $SU(3)$ subgroup.

%Of course only three generations of quarks have been found and they are each
%accompanied by a lepton generation so it seems at first that this model is
%already in conflict with experiment. However this is not quite true since
%the detection of the top quark has still to be confirmed and so, although
%another generation of leptons doesn't seem likely, there is no experimental
%evidence against another generation of quarks.
%In fact,
As stated in section~\ref{expmass}, we take the limits on the masses of a
fourth generation of quarks to be
$M_{b'} > 130$ GeV, $M_{t'}>130$ GeV and the top quark mass to be
$M_t \sim 170$ GeV. We can now use the RGE
equations, first to show that these additional fermions do not cause any
inconsistencies such as gauge coupling constants becoming infinite below the
Planck scale, and then to estimate upper limits on the values of the Yukawa
couplings to the SM Higgs field of these fermions. This will lead to upper
limits on the masses, indicating that the $t'$ and $b'$ quarks would be almost
within reach of present experiments.

\subsection{No Landau Poles}

As in section~\ref{Only5sFails} we can investigate how the gauge coupling
constants vary with energy up to the Planck scale. Here we set the thresholds
for all the unknown fermions (4th generation quarks and fermions coupling to
$SU(5)$), as well as for the top quark and Higgs boson, to $M_Z$. The absence
of Landau poles in this case will guarantee their absence if some of the
thresholds are set higher than $M_Z$. From experimental limits we would expect
that all these thresholds should be greater than $M_Z$.

\begin{figure}
\epsfxsize=\textwidth
\epsffile[150 100 500 500]{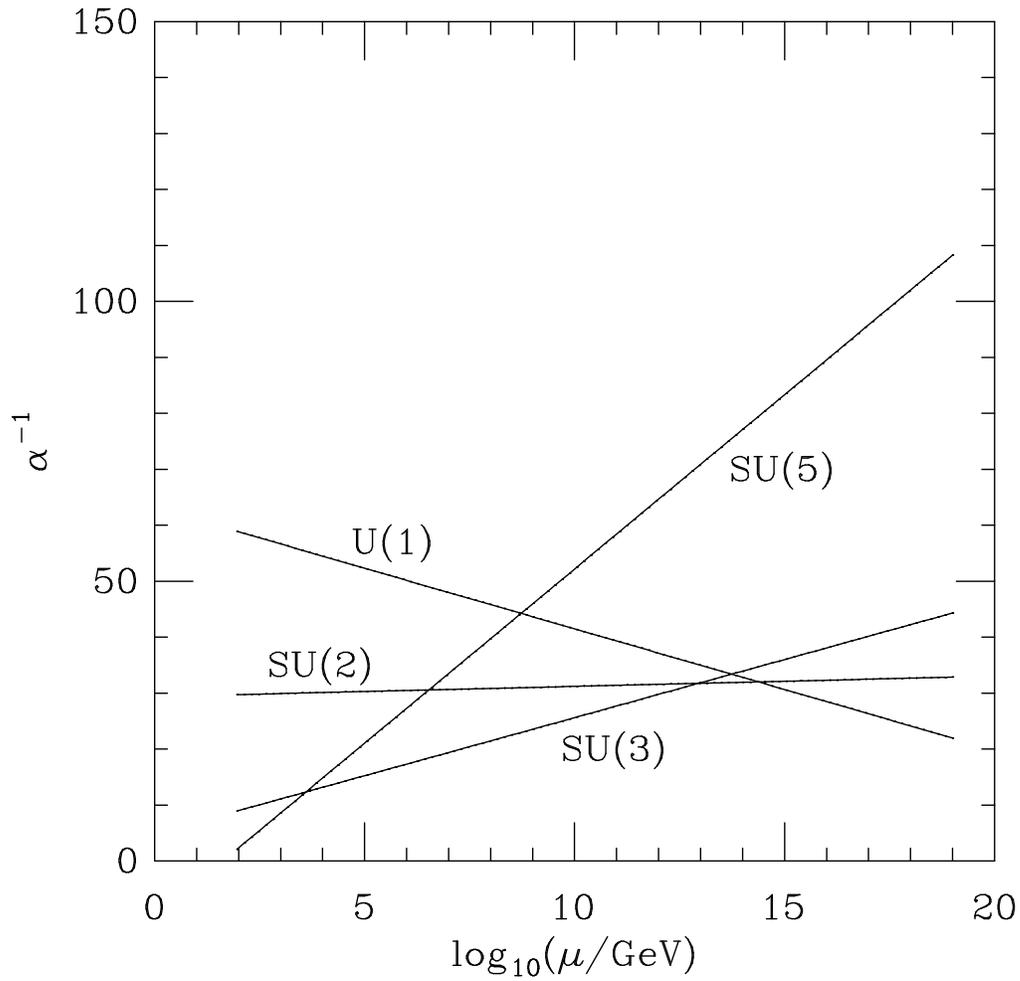}
\caption{$\alpha^{-1}$ from $M_Z$ to the Planck scale for each component group
in the $SMG_{235}$ model with a fourth generation of quarks and the fermions of
table~\protect\ref{SU5Quarks} which couple to $SU(5)$. The initial
value for $\alpha^{-1}_5(M_Z)=2$ was chosen so that it would be confine at the
electroweak scale. There are obviously no Landau poles so this model is
self-consistent.}
\label{alph4g}
%\special{psfile=alph4g.ps angle=270 }
\end{figure}

We use eqs.~(\ref{alpha2})-(\ref{alpha5}) and (\ref{runalpha1GUT}) to run
the gauge coupling constants up to the Planck scale as shown in
Fig.~\ref{alph4g}.
%
%****************
%****************
%
Now we see that with a fourth generation of quarks and the fermions in
Table~\ref{SU5Quarks} (i.e.\ far fewer fermions than the model in
section~\ref{5sOnlySoln} where all the new fermions coupled to $SU(5)$) there
are no problems with Landau poles
below the Planck scale. So our $SMG_{235}$ model with new SM fermions appears
to be consistent.

\subsection{Upper Limits for Yukawa Couplings}
\label{MaxYs}

Now we can choose initial values for the Yukawa couplings at the Planck scale
and use the RGEs to see how they evolve, as they are run down to the
electro-weak scale. Assuming no mixing for the quarks and neglecting the masses
of all SM fermions except the top quark (a good approximation), the RGEs are,
to one loop order in perturbation theory \cite{RGEs1}:
\begin{eqnarray}
\frac{dy_t}{dt} & = & y_t\frac{1}{16\pi^2}
\left(\frac{3}{2}y_t^2+Y_2(S)-G_{3u}\right) \\
\frac{dy_{t'}}{dt} & = & y_{t'}\frac{1}{16\pi^2}
\left(\frac{3}{2}(y_{t'}^2-y_{b'}^2)+Y_2(S)-G_{3u}\right) \\
\frac{dy_{b'}}{dt} & = & y_{b'}\frac{1}{16\pi^2}
\left(\frac{3}{2}(y_{b'}^2-y_{t'}^2)+Y_2(S)-G_{3d}\right) \\
\frac{dy_{5u}}{dt} & = & y_{5u}\frac{1}{16\pi^2}
\left(\frac{3}{2}(y_{5u}^2-y_{5d}^2)+Y_2(S)-G_{5u}\right) \\
\frac{dy_{5d}}{dt} & = & y_{5d}\frac{1}{16\pi^2}
\left(\frac{3}{2}(y_{5d}^2-y_{5u}^2)+Y_2(S)-G_{5d}\right)
\end{eqnarray}
where the $SU(5)$ fermions have been labelled 5u and 5d as generalisations of
the naming of $SU(3)$ quarks. The other variables are defined as
\begin{eqnarray}
Y_2(S) & = & 5y_{5u}^2+5y_{5d}^2+3y_{t'}^2+3y_{b'}^2+3y_t^2 \\
G_{3u} & = & \frac{17}{20}g_1^2+\frac{9}{4}g_2^2+8g_3^2 \\
G_{3d} & = & \frac{1}{4}g_1^2+\frac{9}{4}g_2^2+8g_3^2 \\
G_{5u} & = & \frac{153}{500}g_1^2+\frac{9}{4}g_2^2+\frac{72}{5}g_5^2 \\
G_{5d} & = & \frac{333}{500}g_1^2+\frac{9}{4}g_2^2+\frac{72}{5}g_5^2
\end{eqnarray}
Here $Y_2(S)$ is really $Tr(Y^{\dagger}Y)$ where $Y$ is the Yukawa matrix for
all the fermions.
%The form used above is obviously a good approximation where we
%have neglected all Yukawa couplings much less than 1.

%*****************
\begin{figure}
\epsfxsize=\textwidth
\epsffile[100 100 500 500]{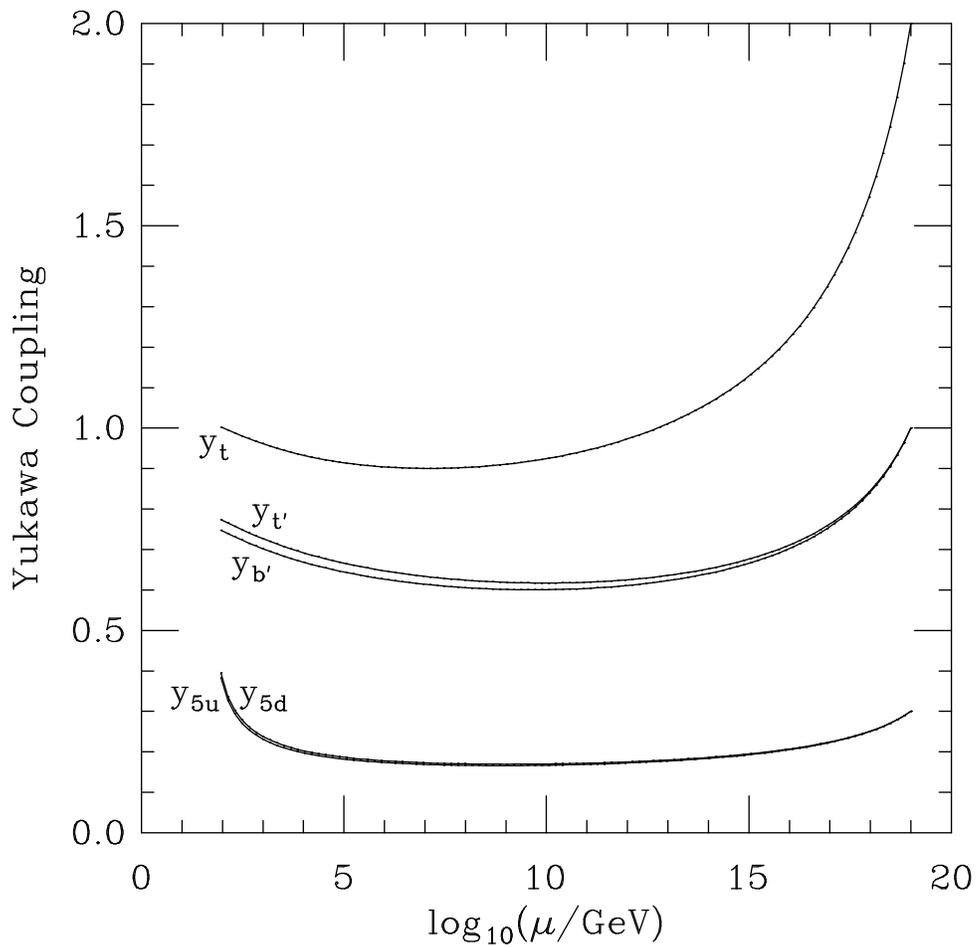}
\caption{An example of running Yukawa couplings for all fermions with
a mass the same order of magnitude as the electroweak scale. The values were
chosen at the Planck scale and run down to $M_Z$ so that all the fermions would
have a mass allowed by current experimental limits.}
\label {yuk4g}
\end{figure}
%*****************

We can choose values for the Yukawa couplings at the Planck scale and then use
the RGEs to see what values the Yukawa couplings will have at any other scale.
We have chosen the low energy scale to be $M_Z$ as shown in Fig.~\ref{yuk4g}.
We observe quasi-fixed points similar to the case for the top quark in the SM
\cite{QuasiFixedPts} and these will provide upper limits on the fermion masses.
However, the resulting Yukawa coupling for any fermion
at $M_Z$ depends on the Yukawa couplings of the other fermions.
Nonetheless there
is an approximate infrared fixed point limit on $Y_2(S)$ and so one Yukawa
coupling can be increased at the expense of the others. This limit on $Y_2(S)$
is quite precise if there is only one strong interaction at low energies, such
as QCD in the SM
\footnote{Detailed results for a general number of heavy SM generations are
derived in \cite{SumM2}.}.
We observe numerically that $Y_2(S) \approx 7.5 \pm 0.3$, provided the Yukawa
couplings of the three heavy quarks are greater than 1 at the Planck
scale and that the Yukawa couplings of the fermions coupling to the $SU(5)$
gauge group are less than the Yukawa couplings of the heavy quarks at
the Planck scale.

The values chosen for Fig.~\ref{yuk4g} have been chosen so that the top quark
pole mass $M_{t} \approx 170 \; \rm{GeV}$ and the fourth generation quark pole
masses are above the current experimental limit of 130 GeV. Also
$M_{b'} \sim M_{t'}$ and $M_{5u} \sim M_{5d}$ have been chosen,
so that there is
only a small contribution to the $\rho$ parameter described in
section~\ref{Epsilons}. We discuss the electroweak
radiative corrections in \cite{4GenLet}
and this $SMG_{235}$ model, with its 8 new doublets, is
consistent with the experimental data at the 2-3 standard
deviation level.
However it is clear that any
model with significantly more $SU(2)$ doublets must disagree with the current
experimental evidence. This rules out the similar models with gauge group
$SMG_{23N}$ where $N$ is an odd integer greater than 5 and not divisible
by 3.

%We discuss the radiative corrections in \cite{4GenLet}
%and this model appears to be consistent with current
%experimental data. However it is difficult to
%calculate the theoretical contributions and the perturbative
%estimates used may not be very accurate. Nevertheless, it is clear that any
%model with significantly more $SU(2)$ doublets must disagree with the current
%experimental evidence. This rules out the similar models with gauge group
%$SMG_{23N}$ where $N$ is greater than 5.

Table~\ref{MaxMasses} gives the values of the Yukawa couplings at $M_Z$ and
the corresponding pole masses, using eq.~(\ref{PoleYuk}), for the quarks. For
the fermions coupling to $SU(5)$ we use the equation relating the pole and
running masses,
\begin{equation}
M_f=\left(1+\frac{12 \alpha_5(M_f)}{5 \pi}\right)m_f(M_f)
\end{equation}

Therefore these masses should be considered upper limits on the masses
of the fermions for this particular choice of Yukawa couplings at the Planck
scale. For other choices of Yukawa couplings at the Planck scale we could, for
example,
increase the mass of the fourth generation of quarks but this would have to be
compensated for by a reduction in the mass of some of the other fermions.
\begin{table}
\caption{Infrared fixed point Yukawa couplings and corresponding pole masses
(for $F_{\pi}=75$ GeV) for a
particular choice of Yukawa couplings at the Planck scale.}
\begin{displaymath}
\begin{array}{|c|c|c|} \hline
$Fermion$ & $Yukawa Coupling at $ M_Z & $Pole Mass (GeV)$ \\ \hline
y_t     & 1.00 & 175    \\ \hline
y_{t'}  & 0.77 & 135    \\ \hline
y_{b'}  & 0.75 & 131    \\ \hline
y_{5u}  & 0.38 & 94    \\ \hline
y_{5d}  & 0.40 & 97    \\ \hline
\end{array}
\end{displaymath}
\label{MaxMasses}
\end{table}

These values for the masses are consistent with current experimental limits but
are not so high that the new fermions could remain undetected for long. In fact
the quark masses may even be within the limits of current accelerators.
It is not clear whether the fermions coupling to $SU(5)$ could be observed,
since they would obviously be confined by the $SU(5)$ gauge interaction which
we take to confine at the electroweak scale. So even if
they have masses of about 100 GeV, they would be much more difficult to detect
than quarks with greater masses. For this reason we consider the clearest
evidence for this model would come from the detection of a fourth generation
quark. The masses of some of the new fermions could be increased, but not by
much, since this would mean a reduction in the mass of other fermions. This
means that this model is consistent and relatively easy to test.

%**********************
\begin{figure}
\epsfxsize=\textwidth
\epsffile[100 100 500 500]{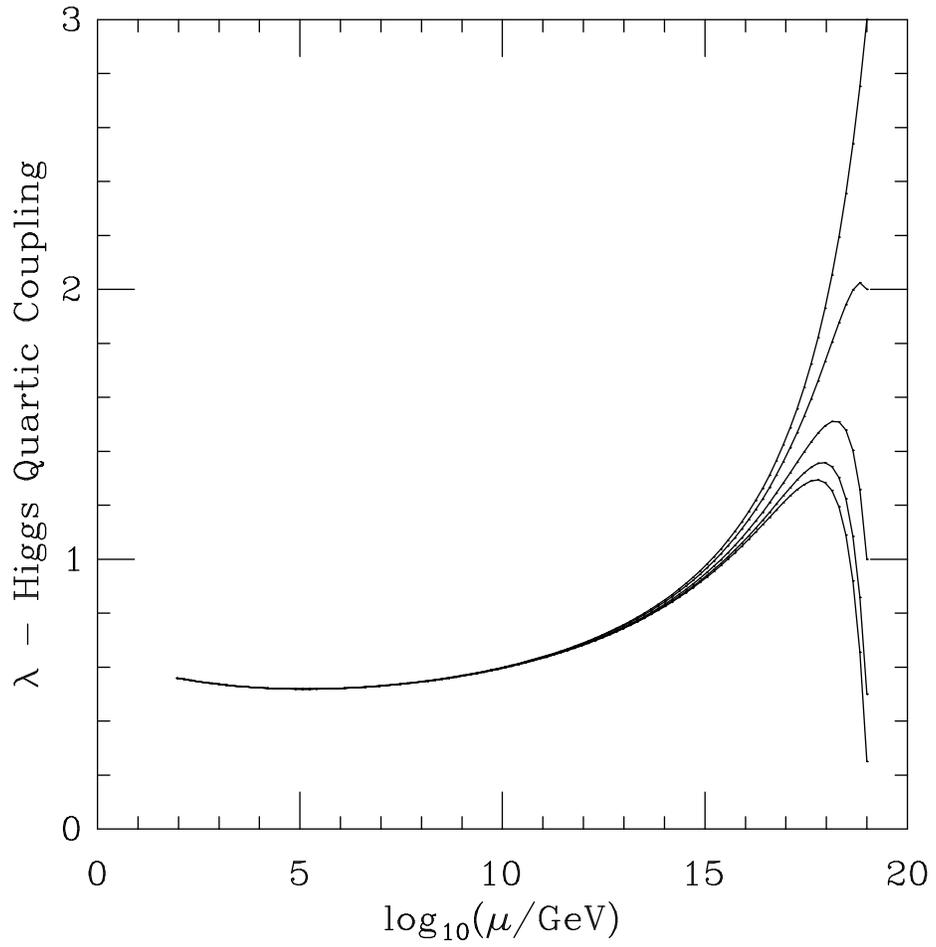}
\caption{Fixed point value of $\lambda$. This graph, along with the estimated
value of $<\phi_{WS}>=234$ GeV, leads to an approximate Higgs mass of 172 GeV.}
\label{lambda}
\end{figure}
%**********************

For completeness we show the running of $\lambda$, the Higgs quartic coupling.
The equation for the running of $\lambda$ is given by \cite{RGEs1},
\begin{eqnarray}
\frac{d\lambda}{dt} & = & \frac{1}{16\pi^2}\left[ 12\lambda^2-\left(
\frac{9}{5}g_1^2+9g_2^2\right) \lambda+ \right. \nonumber \\
 & & \left. \frac{9}{4}\left( \frac{3}{25}g_1^4+\frac{2}{5}g_1^2g_2^2+g_2^4
\right) +4Y_2(S)\lambda-4H(S) \right]
\end{eqnarray}
where we have defined
\begin{equation}
H(S) = 5y_{5u}^4+5y_{5d}^4+3y_{t'}^4+3y_{b'}^4+3y_t^4
\end{equation}
{}From Fig.~\ref{lambda} we obtain $\lambda(M_Z)=0.54$. This graph leads to a
running Higgs mass of
\begin{equation}
M_H(M_H)=\sqrt{\lambda}<\phi_{WS}> \approx 172 \; {\rm GeV}
\end{equation}
The same low energy value of $\lambda$ is obtained for any initial choice of
$\lambda$ at the Planck scale since the Yukawa couplings are at the fixed
point. This means that the Higgs must have this fixed point mass.
If the Yukawa
couplings were slightly lower than their fixed point values we would obtain
a small range of allowable Higgs masses. However, this range would always be
somewhat below
172 GeV.

\section{Conclusions}
\label{Conclusion}
%********************
%\input{conclusion}
We have discussed extensions of the SM having a similar gauge group structure
to the SM itself. In particular we have been guided by the requirement of an
anomaly-free theory, with additional mass protected fermions satisfying a
generalised charge quantisation rule. We were thereby lead to extend the SM
cross product group, $U(1) \otimes SU(2) \otimes SU(3)$, by adding extra
$SU(N)$ direct factors, with the `$N$'s greater than 3 and mutually prime. A
generalised charge quantisation rule, involving each direct factor, was then
obtained by dividing out an appropriate discrete group. Extending the SM in
this fairly obvious way produces the groups $SMG_{23N}$, $SMG_{23MN}$ etc.
Another feature we take over from the SM is the principle of using only small
(fundamental or singlet) fermion non-abelian representations. For the abelian
representations we take the condition that weak hypercharges should be chosen
to be close to zero. More precisely, we minimise the sum of weak hypercharges
squared over all the fermions.

The extra $SU(N)$ groups introduced confine and form fermion condensates having
the same quantum numbers as the SM Higgs doublet. It follows that the extra
$SU(N)$ groups act as partial technicolour groups and must confine near the
electroweak scale. However, the SM Higgs field is still responsible for all the
fermion masses, albeit with a somewhat reduced VEV.

We have studied in detail the conditions for anomaly cancellation in our
minimal extension of the SM gauge group, $SMG_{235}$. It is not possible to
construct
an anomaly-free model using new mass protected fermions which are all
non-singlet under $SU(5)$, without encountering a Landau pole in the $U(1)$
fine structure constant well below the Planck scale. However it is possible to
construct a consistent model with a fourth generation of quarks but, instead of
an extra generation of leptons, with a generation of the fermions coupling to
$SU(5)$ as given in Table~\ref{SU5Quarks}.

A similar solution with a fourth generation of quarks and a generation of
$SU(N)$ fermions as given in Table~\ref{SUNQuarks} is possible for the gauge
group $SMG_{23N}$. However the number of $SU(2)$ doublets in the model
increases with $N$ and hence their contribution to the electroweak radiative
corrections becomes more important. The $SMG_{235}$ model is just consistent
with the precision electroweak data but $SMG_{23N}$ models with $N>5$ are
probably ruled out \cite{4GenLet}. Similarly the $SMG_{23MN}$ models would be
inconsistent with the precision electroweak data.

The $SMG_{235}$ model with a fourth generation of quarks and a generation of
$SU(5)$ fermions seems to be phenomenologically consistent. It requires the
existence of $t'$ and $b'$ quarks at or below the top quark mass scale; this is
consistent with current experimental limits but they could not remain
undetected for long. However it is unlikely that the $SU(5)$ fermions could be
observed with current accelerators; they would be confined inside $SU(5)$
`hadrons', with a confinement scale of order 200 GeV and would have a small
production cross section at present hadron colliders. Even if this
model doesn't turn out to be correct we hope that the derivation might at
least highlight some of the important features of the SM and some of the unique
qualities of the SM, which appears (admittedly almost by definition) as the
smallest case of our more general models.

%********************
%
%********************
%\input{acknowl}
\vspace{1cm}

\centerline{\bf Acknowledgements}
\vspace{.3cm}

We thank John Gunion, Hans Jensen, Victor Novikov, David Sutherland and Misha
Vysotsky for helpful discussions. This work has been supported in part by
INTAS Grant No.\ 93-3316 and PPARC Grant No.\ GR/J21231, the British Council,
Cernf{\o}lgeforskning and EF contract SC1 0340 (TSTS).

%********************
%
%
\newpage
\appendix
\section{Deriving the SM Generation}
\label{SMGen}
%************
%\input{smgen}
\subsection{The SM Generation}

In the SM there are 3 generations of fermions which are identical except for
their masses. Each generation consists of 15 Weyl fermions and can be divided
into a lepton generation and a quark generation. The quarks couple to the
$SU(3)$ gauge group, whereas the leptons are $SU(3)$ singlets and so do not
`feel' the strong force. The properties of these fermions are shown in
table~\ref{SMGeneration}.
The fermions are labelled as in the first (lightest) generation.
\begin{table}[bht]
\caption{The lightest SM generation.}
\begin{displaymath}
\begin{array}{|c|c|c|c|c|}
\hline
$Generation$ & $Fermion$ & $Representation of$ & \rm{Representation} &
						\rm{Electric\;Charge} \\
& $Label$ & SU(2) \otimes SU(3) & {\rm of\;} U(1) {\rm , \;} \frac{y}{2} & Q \\
\hline
	& \left( \begin{array}{c} u \\ d \end{array} \right)_{L} &
{\bf 2},{\bf 3} & \frac{1}{6} & \left( \begin{array}{r} \frac{2}{3} \\
-\frac{1}{3}
				\end{array} \right) \\[15pt] \cline{2-5}
$Quark$ & \overline{u}_{L} & {\bf 1},{\bf \overline{3}} & -\frac{2}{3} &
-\frac{2}{3} \\ \cline{2-5}
	& \overline{d}_{L} & {\bf 1},{\bf \overline{3}} & \frac{1}{3} &
\frac{1}{3} \\ \hline
$Lepton$ & \left( \begin{array}{c} \nu_{e} \\ e \end{array} \right)_{L} &
{\bf 2},{\bf 1} & -\frac{1}{2} & \left( \begin{array}{r} 0 \\
			-1 \end{array} \right) \\[15pt] \cline{2-5}
	& \overline{e}_{L} & {\bf 1},{\bf 1} & 1 & 1 \\ \hline
\end{array}
\end{displaymath}
\label{SMGeneration}
\end{table}

The quark generation is formed by the representations
$(\frac{1}{3},{\bf 2},{\bf 3})_L$,
$(-\frac{4}{3},{\bf 1},{\bf \overline{3}})_L$ and $(\frac{2}{3},{\bf 1},{\bf
\overline{3}})_L$ of the
gauge group $U(1) \otimes SU(2) \otimes SU(3)$. This is precisely the mass
grouping $\{\frac{1}{3},{\bf 3}\}$ (where
the representation {\bf 3} is of the gauge group $SU(3)$) described in
section~\ref{Simplifications}. All the quarks get a mass by the Higgs
mechanism. The lepton generation is formed by the representations
$(-1,{\bf 2},{\bf 1})_L$ and $(2,{\bf 1},{\bf 1})_L$ of the same gauge group.
However, this is
not the same as the mass grouping $\{-1,{\bf 1}\}$ because there is no
right-handed
neutrino (representation $(0,{\bf 1},{\bf 1})_L$) in the SM\@. This means that
the neutrino is massless in the SM but the
electron can still get a mass by the Higgs mechanism. However, the
lepton generation gives the same contribution to all anomalies as the mass
grouping $\{-1,{\bf 1}\}$ would, since the right-handed neutrino would be
totally
neutral (i.e.\ would not interact with any gauge fields).

%*****************
%\input{miny2}
%*****************
\subsection{Derivation of the SM Generation}
\label{DerSMGen}

In fact, we can derive the SM generation
using the following assumptions \cite{OrigOfSym}: \\
(i) {\em The SM gauge group}: $SMG \equiv S(U(2)\otimes U(3))$. This includes
the charge quantisation rule eq.~(\ref{SMchqu}). \\
(ii) {\em Mass protection}: This means that no fermions can form a gauge
invariant mass term except by the Higgs mechanism. In particular we cannot have
left and right handed fermions with the same representation of the $SMG$. Also
we cannot have a right handed neutrino since it can get a Majorana mass. \\
(iii) {\em Anomaly cancellation}: In addition to the cancellation of gauge
anomalies, the Witten global $SU(2)$ anomaly and the mixed gauge and
gravitational anomaly must also be absent. \\
(iv) {\em Small representations}: This means (c.f.\ section~\ref{SmallReps})
that all fermions are in either fundamental
or singlet representations of the $SU(2)$ and $SU(3)$ subgroups and the sum
of weak hypercharge squared for all fermions is as small as possible.

So our aim is to minimise the value of
$\sum_{i} S_{i} \left ( \frac{y_{i}}{2} \right )^{2}$ (where $S_i$ is the
dimension of representation $i$ with weak hypercharge $y_i$) for all possible
choices of mass protected fermions in fundamental or singlet
representations of $SU(2)$ and $SU(3)$, assuming the charge quantisation rule,
eq.~(\ref{SMchqu}), and cancelling all relevant anomalies. We note that for
one SM generation (which satisfies assumptions (i) to (iii))
\begin{equation}
\label{SMy2}
\sum_{i} S_{i} \left ( \frac{y_{i}}{2} \right ) ^{2} = \frac{10}{3}
\end{equation}
and we show that there is no other mass protected solution of the anomaly
constraints with
\begin{equation}
\label{y2limit}
\sum_{i} S_{i} \left ( \frac{y_{i}}{2} \right ) ^{2} \le \frac{10}{3}
\end{equation}
So we shall prove that one SM generation also satisfies assumption (iv) and
thus we will show that assumptions (i) to (iv) define the SM generation. Note
that in order to satisfy assumption (iv) we must satisfy eq.~(\ref{y2limit}).
So in the following analysis we will implicitly assume eq.~(\ref{y2limit}).
Table~\ref{SMTypes} shows all allowed representations and their contribution
of $S \left ( \frac{y}{2} \right ) ^{2}$.
\begin{table}
\caption{Contributions of $S \left ( \frac{y}{2} \right )^2$ for all
fundamental
and singlet representations of $SU(2)$ and $SU(3)$ for any value of weak
hypercharge which satisfies eq.~(\protect\ref{SMchqu}\protect). All `$N$'s are
integers and S is the dimension of the non-abelian representation.}
\begin{displaymath}
\begin{array}{|c|c|c|c|} \hline
 & {\rm Representation \; of} & & \\
{\rm Type} & SU(2) \otimes SU(3) & \frac{y}{2} & S \left ( \frac{y}{2} \right )
^{2} \\ \hline
a & {\bf 2},{\bf 3} & N_a+\frac{1}{6} & 6N_a^2+2N_a+\frac{1}{6} \\ \hline
b & {\bf 2},{\bf \overline{3}} & N_b-\frac{1}{6} & 6N_b^2-2N_b+\frac{1}{6} \\
\hline
c & {\bf 1},{\bf 3} & N_c-\frac{1}{3} & 3N_c^2-2N_c+\frac{1}{3} \\ \hline
d & {\bf 1},{\bf \overline{3}} & N_d+\frac{1}{3} & 3N_d^2+2N_d+\frac{1}{3} \\
\hline
e & {\bf 2},{\bf 1} & N_e-\frac{1}{2} & 2N_e^2-2N_e+\frac{1}{2} \\ \hline
f & {\bf 1},{\bf 1} & N_f & N_f^2 \\ \hline
\end{array}
\end{displaymath}
\label{SMTypes}
\end{table}

In order to satisfy eq.~(\ref{y2limit}) we must choose $N_a=N_b=0$,
$N_c \in \{0,1\}$, $N_d \in \{-1,0\}$, $N_e \in \{0,1\}$ and
$N_f \in \{-1,1\}$. (We don't consider $N_f=0$ because this would be a
right-handed neutrino which would not contribute to any anomalies and would be
expected to get a Majorana mass of the order of the Planck mass). This means
that we cannot have mass protected fermions of types $a$ and $b$. So we can
choose, without loss of generality, that there are no fermions of type $b$
\footnote{Choosing no fermions of type $a$ would lead to an equivalent solution
with opposite chirality.}.
So we get Table~\ref{AllowedTypes}, which shows all
allowed fermions and contributions to some anomalies.

\begin{table}
\caption{All allowed representations of fermions which could be used to satisfy
eq.~(\protect\ref{y2limit}\protect) and their contributions to some anomalies.}
\begin{displaymath}
\begin{array}{|c|c|c|c|c|c|c|} \hline
 & \rm{Representation \; of} & & & & & \\
{\rm Type} & SU(2) \otimes SU(3) & \frac{y}{2} & S \left ( \frac{y}{2} \right )
^2 &  [SU(3)]^3 & [SU(3)]^2U(1) & [SU(2)]^2U(1) \\ \hline
a & {\bf 2},{\bf 3} & \frac{1}{6} & \frac{1}{6} & 2 & \frac{1}{3} & \frac{1}{2}
\\ \hline
c_1 & {\bf 1},{\bf 3} & -\frac{1}{3} & \frac{1}{3} & 1 & -\frac{1}{3} & 0 \\
\hline
c_2 & {\bf 1},{\bf 3} & \frac{2}{3} & \frac{4}{3} & 1 & \frac{2}{3} & 0 \\
\hline
d_1 & {\bf 1},{\bf \overline{3}} & \frac{1}{3} & \frac{1}{3} & -1 & \frac{1}{3}
& 0 \\ \hline
d_2 & {\bf 1},{\bf \overline{3}} & -\frac{2}{3} & \frac{4}{3} & -1 &
-\frac{2}{3} & 0 \\ \hline
e_1 & {\bf 2},{\bf 1} & -\frac{1}{2} & \frac{1}{2} & 0 & 0 & -\frac{1}{2} \\
\hline
e_2 & {\bf 2},{\bf 1} & \frac{1}{2} & \frac{1}{2} & 0 & 0 & \frac{1}{2} \\
\hline
f_1 & {\bf 1},{\bf 1} & -1 & 1 & 0 & 0 & 0 \\ \hline
f_2 & {\bf 1},{\bf 1} & 1 & 1 & 0 & 0 & 0 \\ \hline
\end{array}
\end{displaymath}
\label{AllowedTypes}
\end{table}

For mass protection we cannot have any of the following combinations; types
$c_1$ and $d_1$, types $c_2$ and $d_2$, types $e_1$ and $e_2$, or types $f_1$
and $f_2$ (all defined in Table~\ref{AllowedTypes}). Also note that all the
types of representations in Table~\ref{AllowedTypes} contribute to the mixed
anomaly, $\sum_i S_iy_i$. This means that we cannot use only type $f$ fermions
to produce an anomaly-free set of mass protected fermions. Therefore, if no
fermions couple to the $SU(3)$ group, there is no way to cancel the
$[SU(2)]^2U(1)$ anomaly. So we can conclude that some fermions must couple to
$SU(3)$.

Suppose there are no fermions of type $a$. Then the above arguments mean that,
to cancel the $[SU(3)]^3$ anomaly, we must have equal numbers of either types
$c_1$ and $d_2$ or types $c_2$ and $d_1$. But then there is no way to cancel
the $[SU(3)]^2U(1)$ anomaly. So we have a contradiction, which means that there
must be at least one type $a$.

The $[SU(2)]^2U(1)$ anomaly must be cancelled by having as many type $e_1$ as
type $a$. So there are no type $e_2$ due to the principle of mass protection.
Again using the principle of mass protection, the only way to cancel the
$[SU(3)]^3$ and $[SU(3)]^2U(1)$ anomalies is by having the number of types
$a$, $d_1$ and $d_2$ the same. We can now cancel the $[U(1)]^3$ and mixed
anomalies using Table~\ref{FinalTypes}.

\begin{table}
\caption{Allowed combinations of fermions and their contribution to the
remaining anomalies.}
\begin{displaymath}
\begin{array}{|c|c|c|c|} \hline
\rm{Types} & G^2U(1) & [U(1)]^3 & S \left ( \frac{y}{2} \right )^2 \\ \hline
a+d_1+d_2+e_1 & 1+1-2-1=-1 & \frac{1}{36}+\frac{1}{9}-\frac{8}{9}-\frac{1}{4}
=-1 & \frac{7}{3} \\ \hline
f_1 & -1 & -1 & 1 \\ \hline
f_2 & 1 & 1 & 1 \\ \hline
\end{array}
\end{displaymath}
\label{FinalTypes}
\end{table}

So we see that the anomaly-free set of mass-protected fermions which
minimises the sum of the weak hypercharges squared, is one of type $a$, $d_1$,
$d_2$, $e_1$ and $f_2$. This is one SM quark-lepton generation.

%*****************
%\input{altSMGen}
%*****************
\subsection{Alternative Derivations of the SM Generation}

There have been other attempts to derive the SM generation using various
assumptions. Most notably Geng and Marshak \cite{Marshak} have tried to derive
the SM generation using the constraints due to cancellation of anomalies. They
also assume mass protection but not the charge quantisation rule
eq.~(\ref{SMchqu}). Instead of minimising the sum of weak hypercharges squared,
they try to find the minimum number of fermions required to satisfy these
assumptions.

The smallest number of Weyl fermions found by Geng and Marshak is 14. This
solution consists of the following representations of the gauge group
$U(1) \otimes SU(2) \otimes SU(3)$: $(0,{\bf 2},{\bf 3})_L$, $(y,{\bf 1},{\bf
\overline{3}})_L$,
$(-y,{\bf 1},{\bf \overline{3}})_L$ and $(0,{\bf 2},{\bf 1})_L$. They rule out
this solution
because the $SU(2)$ doublet cannot acquire a Dirac or Majorana mass, even with
the spontaneous symmetry breaking of the gauge group. However, we know from
the SM that the neutrino is massless and so there doesn't appear to be any
reason why massless fermions should be excluded from such an analysis. (We
could obviously use phenomenological arguments but that would defeat the
purpose of trying to derive the SM generation). They also object to this
solution because they feel it trivialises the cancellation of the mixed
gravitational and gauge anomaly. In what sense the anomaly condition is trivial
is not entirely clear, since not all fermions have zero weak hypercharge; but
also why should it matter if a constraint is trivially satisfied? In our
derivation this solution
does not occur because of the charge quantisation rule. So by enforcing the
charge quantisation rule, which we have taken as one of the defining properties
of the $SMG$ in section~\ref{Groups}, we can avoid this solution without
introducing dubious arguments about fermion masses.

%So, if we add the assumption of the charge quantisation rule, we would expect
%to find that the SM generation is the smallest possible number of Weyl
%%fermions.
%However, there are two smaller solutions which have not been considered by
%%Geng
%and Marshak. The smallest is made of the representations; $(y_1,2,1)_L$,
%$(y_2,2,1)_L$, $(y_3,2,1)_L$ and $(y_4,2,1)_L$, a total of 8 Weyl fermions.
%The other consists of the representation; $(y_1,1,3)_L$, $(y_1,1,3)_L$,
%$(y_1,1,\overline{3})_L$ and $(y_1,1,\overline{3})_L$, a total of 12 Weyl
%fermions. In both cases the values of weak hypercharge, $y_i$, can be chosen
%so that $\sum_{i=1}^{4}y_{i}=0$ for anomaly cancellation and also
%so that all fermions are mass protected. These solutions appear to have
%been ignored because Geng and Marshak implicitly assumed that at least one
%fermion must couple to each part of the gauge group. Obviously the above
%solutions do not couple to the $SU(3)$ and $SU(2)$ subgroups respectively.

If we then also add the assumption that all subgroups must have some fermion
coupling to them, we can almost derive the SM generation. The problem is that
we can scale all values of weak hypercharge for the SM fermions by a factor of
$(6n+1)$ where $n$ is any integer
\footnote{Without the charge quantisation rule we could scale the weak
hypercharges by an arbitrary amount. Then we couldn't use the procedure of
minimising the sum of hypercharges squared, since this would obviously force
all values to zero. There is then no way to fix the scale other than by
assuming the fermions get a mass by the Higgs mechanism and fixing the scale
to the weak hypercharge of the Higgs boson. So the charge quantisation rule
effectively introduces a scale for the weak hypercharge independent of any
Higgs bosons.}.
The SM generation is obviously the solution
with the values of hypercharge closest to zero. We can express this by choosing
to minimise the sum of hypercharges squared for this solution. But since we
must introduce such an assumption why not use it from the start?! This then
allows us to drop two of the above assumptions; that all subgroups must have
a fermion coupling to them and that we should look for the smallest number of
Weyl fermions. We are then left with the four assumptions used in
section~\ref{DerSMGen}.
This seems
more reasonable than introducing more assumptions with no justification.

%************
%
%************
%\include{bib}

%************
%
\end{document}